\documentclass[namedreferences]{solarphysics}
\usepackage[hyperref,optionalrh,solaromanenum]{spr-sola-addons} % For Solar Physics
\usepackage{epsfig}                     % For eps figures, old commands
\usepackage{graphicx}        % For eps figures, newer & more powerfull
\usepackage{color}                       % For color text: \color command
\usepackage{breakurl}                         % For breaking URLs easily trough lines
\begin{document}

\begin{article}

\begin{opening}

\title{Numerical MHD Simulation of the Coupled Evolution of
Collisional Plasma and Magnetic Field in the Solar
Chromosphere. I. Gradual and Impulsive Energisation\\ {\it Solar Physics}}

%%%%%%%%%%%%%%%%%%%%%%%%%%%%%%%%%%%%%%%%%%%%%%%%%%%
%% Authors Names
%

\author[addressref={lma},corref,email={l.m.alekseeva@yandex.ru}]{\inits{L.M. }\fnm{L.M. }\lnm{Alekseeva}}
\author[addressref={spk},email={renger@mail.ru}]{\inits{S.P. }\fnm{S.P.~}\lnm{Kshevetskii}}

%%%%%%%%%%%%%%%%%%%%%%%%%%%%%%%%%%%%%%%%%%%%%%%%%%%
%% Runningheads
%
\runningauthor{L.M. Alekseeva, S.P. Kshevetskii} \runningtitle{MHD
Simulation of Chromospheric Plasma}

%%%%%%%%%%%%%%%%%%%%%%%%%%%%%%%%%%%%%%%%%%%%%%%%%%%
%% Affilations
%% id shold be the same with \author addressref value.
\address[id={lma}]{Skobeltsyn Institute of Nuclear Physics, Lomonosov Moscow State
University, Moscow, 119991 Russia}
\address[id={spk}]{Immanuel Kant Baltic Federal University,
Kaliningrad, 236041 Russia}

%%%%%%%%%%%%%%%%%%%%%%%%%%%%%%%%%%%%%%%%%%%%%%%%%%%
%%% Abstract
\begin{abstract}
The dynamical coupling between the solar chromospheric
plasma and magnetic field is investigated by numerically
solving a fully self-consistent, two-dimensional initial-value
problem for the nonlinear collisional MHD equations including
electric resistivity, thermal conduction, and, in some cases, gas-dynamic
viscosity. The processes in the contact zone between two horizontal magnetic
fields of opposite polarities are considered. The plasma is
assumed to be initially motionless and having a temperature of
50,000~K uniform throughout the plasma volume; the characteristic magnetic
field corresponds to a plasma $\beta\gtrsim 1$.
In a physical-time interval of 17~seconds typically covered by a
computational run, the plasma temperature gradually increases
by a factor of two to three. Against this background, an
impulsive  (in 0.1~seconds or less) increase in the current-aligned plasma velocity
occurs at the site of the current-layer thinning (sausage-type
deformation, or $m=0$ pinch instability). Such a ``velocity burst''
 can be interpreted physically as an event of suprathemal-proton generation.
Further development of the sausage instability results in
an increase in the kinetic temperature of the protons to high values, even
to those observed in flares. The form of our system of MHD
equations indicates that such increases are a property of
the exact solution of the system at an appropriate choice of the
parameters. Magnetic reconnection does not manifest itself in this solution: it would
generate flows forbidden by the chosen geometry.
Therefore, the pinch-sausage  effect can act as an energiser
  of the upper chromosphere and be an alternative to the magnetic-reconnection process
  as the producer of flares.
\end{abstract}

%%%%%%%%%%%%%%%%%%%%%%%%%%%%%%%%%%%%%%%%%%%%%%%%%%%
%% Keywords
%
\keywords{Magnetohydrodynamics; Plasma Physics; Magnetic fields; Chromosphere;
Heating, Chromospheric; Transition Region; Jets; Energetic Particles, Protons}

\end{opening}
%-------------------------------------------------

%%%%%%%%%%%%%%%%%%%%%%%%%%%%%%%%%%%%%%%%%%%%%%%%%%%
%% Sections
%
\section{Introduction}\label{intr}
The chromosphere is much less studied than the underlying photosphere and
overlying corona. However, its properties are widely discussed in the context
of the fundamental problems of solar physics \citep{ASCH,Flet}. Much
effort has been made to understand the mechanism of heating of the upper
chromosphere and corona. Such a mechanism is usually associated with coronal
processes at plasma-parameter values $\beta < 1$. On the other hand, some
observed phenomena lead, as Aschwanden notes, ``to a paradigm shift from the
coronal heating problem to a dynamic chromospheric energisation problem''
\citep{ASCH2001}; the latter corresponds to the case of $\beta \geq1$
\citep{ASCHN,ASCH2008}. Therefore, particular attention should be given to
plasma properties at $\beta \approx 1$.

Progress in understanding these issues will remain slow as long
as the knowledge of the chromosphere remains fragmentary. In
many cases, detailed information provided by the modern
high-resolution instruments can hardly be compared with
theoretical conclusions, since  the theories are based on
non-observable parameters (nonpotential magnetic field, magnetic
stress or twist, electric fields, \textit{etc.}) \citep{ASCH2008}.

 Such difficulties in studying the dynamics of the
chromospheric plasma itself could be avoided using the
completely self-consistent system of MHD equations. Various
quantities are mutually related in a solution of this system,
and a passage from some quantity crucial for the process to
another, more convenient for an observer, can be done quite
routinely.

However, in the practice of simulations of the chromospheric plasma, even the
inclusion of magnetic fields increases enormously the level of complexity
\citep{Carl}. Highly dynamic, nonlinear processes are typical of the
chromosphere. Small scales are involved in them due to both turbulent chaos and
the formation of shock fronts. \inlinecite{ASCH} writes that ``typical
difficulties with the MHD method are the same as they are typical for numerical
simulations, such as heavy computing demand, convergence problems, insufficient
spatial resolution to handle discontinuities, line-tying (continuous slippage
of magnetic-field lines)''.

However, direct simulations of dynamical coupling between laboratory
ion--electron plasma and magnetic
 field were made by
numerically solving a two-dimensional initial-value problem for fully
self-consistent, nonlinear collisional MHD equations, which give a unified
description for $\beta > 1$, $\beta \approx 1$, and $\beta < 1$
\citep{BM,Brushl1989}. Later, various situations were considered in the context
of studying laboratory plasma-channel flows
(see an overview by \citep{A}).
It is worthwhile to try to comprehend how the features of the dynamical coevolution of
the plasma and magnetic field, revealed in this way, can manifest themselves
under solar conditions. To this end, we solve here numerically an initial-value
problem of this sort; in so doing, as in \cite{BM}, we do not assume any
properties of the solutions sought for. In view of investigating the
solar-plasma dynamics, we employ a substantially improved numerical technique.
This makes it possible to describe small-scale turbulence and
discontinuities, automatically passing, whenever necessary, to the use of the
class of generalised functions. This technique was verified in our previous
studies \citep{Ksh-Gavr,Ksh06,A-K-1,A-K-SAO}). This allows us to obtain
solutions for comparatively long time intervals even if small-scale
perturbations develop.

We consider a two-dimensional problem assuming that all physical quantities are
constant along the magnetic-field lines and the magnetic field is horizontal.
This is motivated by the fact that about 95\,\% of the magnetic flux issuing
from the photosphere closes below the coronal heights \citep{Priest,ASCH}, so
that the magnetic field in the chromosphere consists largely of horizontal
apical segments of the field lines; horizontal segments are also present in the
field lines of the magnetic canopies.

  We investigate here the coevolution of the collisional
electron--proton plasma and magnetic field at the characteristic plasma
parameter values $\beta_{0*}=1.5$, 1.6, or 2.3, which are present in the upper
chromosphere \citep{Gary} [see also Figure~1.22 in \inlinecite{ASCH}].
 Our MHD simulations are carried out under the assumption that the
plasma is initially motionless and has a temperature of 50,000~K everywhere in
the computation domain.

 Such fully ionized plasma can be attributed to the upper chromosphere. The
upper boundary of the chromospheric layer is typically associated with a
temperature of about 50,000~K; this layer is assumed to be overlaid with a
transition region, which, in turn, changes into the corona at temperatures of
about 500,000~K \citep{Gabriel}. However, more recent observations modified our
understanding of the geometry of the solar atmosphere, which is now regarded as
an inhomogeneous mixing of photospheric, chromospheric, and coronal zones
produced by various dynamic processes \citep{Schrijver} [see also Figure~1.17 of
\inlinecite{ASCH}]. Since we consider small-size regions, there is generally no
need to attribute them to particular layers, and the temperature can be assumed
to be the basic parameter characterising the plasma under study.\footnote{To
avoid misunderstanding, it is worth noting that the definition of the upper
chromosphere used by \inlinecite{Gabriel} is not the one in common usage, and
there are also other different  definitions, usually based on the characteristics
of neutral-particle distributions over the solar atmosphere. We can generally
consider plasma with temperatures of above 20,000~K using the equations for
the fully ionised gas.}

\inlinecite{A-K-SAO} carried out simulations for a characteristic plasma-
$\beta_{0*}=1.5$ and an initial magnetic configuration consisting of two
regions with oppositely directed magnetic fields. We noted the development of a
pinch-type plasma instability. Some its features are also observed in
laboratory plasmas; these are the contraction of the current layer as a whole,
the local thinning (sausage-type deformation) of the stagnated current layer,
and the development of oppositely directed, current-aligned plasma streams.
However, the process in an unbounded solar plasma proved to be more complex.
Another pair of jets accompanied by a strip of small-scale inhomogeneities was
directed transversely to the electrical current. As viewed along the
magnetic-field lines, the pattern of field, velocity, and density distributions
appeared to be cross-shaped; we shall call such a pattern a \textit{transient
``cross-shaped'' structure}. The jet velocity exceeded 20~km\,s$^{-1}$. A class
of situations was identified in which the plasma temperature increased in the
central part of this pattern (the computations were terminated when it
doubled).

Here, we shall demonstrate that impulsive processes can develop against the
background of a gradual development of this pattern (which will be illustrated
in the case of $\beta_{0*}=1.6$). Specifically, intense plasma streams emerge
in a temporal  interval of about 0.1~second. Such a \emph{velocity burst} can physically
be interpreted as an event of suprathermal-ion generation. The kinetic
temperature based on the maximum velocity can reach about 0.5~MK. This is
indicative for the dynamic energisation of a parcel of chromospheric plasma to
coronal temperature. We conclude that the velocity burst is due to a
sausage-type pinch instability. This velocity peak is neither unique nor the
highest one. The very form of the initial-value problem considered here
indicates an important property of its exact solution: the development of the
sausage instability ultimately results in an infinite growth of the plasma
velocity [$v$] with time [$t$], which implies an increase in the kinetic
temperature to high values, even such as observed in flares. Our numerical
solutions allow us to trace this process.

The infinite velocity  growth can naturally be suppressed by gas-dynamic molecular viscosity,
which will be observed in simulations with a strongly exaggerated
viscosity.
 Some other factors may also act
similarly. Analyses of their role fall beyond the scope of our
study. We are interested in investigating the possibility
of impulsive plasma energisation by the pinch instability and,
following \inlinecite{Carl}, ``perform numerical experiments
and modelling in simplified cases in order to fashion a basic
physical foundation upon which to build our understanding.''

\section{Magnetohydrodynamic Background}\label{our}
\subsection{Collisional Plasma and its Dynamical Properties
Depending on the Rarefaction Degree}\label{our1}
 Some particular phenomena known in the theory of
accelerating plasma channels may be of interest to solar physics.
Two-dimensional flows transverse to the magnetic field with the physical
quantities constant along the magnetic-field lines have long been studied.
Early numerical simulations\,--\,generally, with a finite electric conductivity
and the Hall effect taken into account \,--\,discovered the development of
closed solitary electric-current structures, i.e. isolated magnetic tubes
\citep{AS,BGM,MS,BM}. Without the Hall effect, they develop at $\beta \geq1$,
and they do not depend on the presence of walls confining the plasma
\citep{AS,MS,BM}. The role of the Hall effect in the dynamics of a dense plasma
with a magnetic field is determined by the dimensionless parameter
\begin{equation}\label{xi}
\xi=c(eL)^{-1}\sqrt{m_\mathrm i /(4\pi N_*)},
\end{equation}
which will be referred to as the Hall-plasma-dynamics parameter; here, $c$,
\mbox{$m_\mathrm i$}, $L$, and $N_*$ are, respectively, the speed of light, the
proton mass, the length scale, and the reference concentration of particles of
the a given electrical-charge sign. The plasma is assumed to be quasi-neutral,
with a large but finite $N_*$ \citep{BM}. In a Hall plasma, although structures
 of the above-mentioned sort arise at  $\beta >
1$, they manifest themselves more clearly (and with sharper
gradients), as $\beta$ decreases, being related to explosive events
characteristic of Hall plasmas \citep{BM}. These phenomena are, to all
appearance, due to a global gradient of the gas pressure in the flow
direction \citep{AS,BM,AT1,AT2,ABASE}.

In the solar atmosphere, the Boltzmann-distribution-based
pressure gradient can
produce similar phenomena. A stationary two-dimensional
solution of the same equation system was found analytically for
the case of a very weak magnetic field in a Hall plasma with
the presence of gravity \citep{A}. It represents a thin
layer of a current directed upward, with a downward plasma flow
and a slower return flow at the periphery of this layer. It
changes into a thin current sheet in a more rarefied plasma,
which may be of interest in the context of magnetic
reconnection. \inlinecite{G} used a steady-state MHD model of
the collisional chromospheric Hall plasma, with the
electrical-conductivity and thermolectric tensors taken into
account in Ohm's law. For special conditions of a given
magnetic field of a current-layer type and a given bulk plasma
flow orthogonal to the magnetic field, he found that the plasma
thermodynamic parameters in a certain part of the current layer
correspond to the lower-corona range.

These results prompted us to start a systematic numerical investigation based
on the collisional MHD equations for fully ionised chromospheric layers,
generally with the inclusion of gravity, the Hall effect, and time-dependent
electric and thermal conduction. To solve an axisymmetric or planar-symmetric
initial-value problem, we applied a special numerical method \citep{Ksh-Gavr} that
 uses generalised functions wherever necessary, to the solar plasma.

\subsection{MHD Chromospheric Processes in our Two-Dimensional Numerical Simulations}
\label{our2}
The penetration of the magnetic field into the chromosphere
immediately and typically entails i) a tendency of the gas pressure and
magnetic pressure to come into balance and ii) the interaction between the
penetrating magnetic field and other closely located chromospheric magnetic
fields. Generally, the two phenomena parallel each other and both produce wavelike
steadying processes (from here on, referred to as \textit{primary waves}),
which interfere.

We investigate this typical situation for the case of two contacting magnetic
regions of opposite polarities. (For simplicity, we choose the initial
distributions of the variables to be symmetric with respect to the zero-field
surface. We also assume the physical quantities to be constant along the
magnetic-field lines, which are horizontal, straight and parallel; therefore, a
2D problem can be formulated.) In this circumstances, the primary waves
associated above with phenomenon ii) are due to electrical-current contraction
in the plasma medium and are formed by the compression and subsequent expansion
of the whole current zone between the regions. Also, a sausage instability
similar to that in the laboratory Z-pinch can be expected at some locations in
the same zone, in the presence of the primary steadying waves and/or after they
recede.

Our numerical method allows us to calculate inhomogeneities (waves, turbulence,
and shocks) spontaneously arising during the nonlinear process of plasma and
magnetic-field coevolution. We do this without anticipating any properties of
the solution. Thus, solving an initial-value problem for the fully consistent
system of MHD equations, we can perform direct numerical simulations specifying
only the initial conditions.

Our early simulations at $\xi\neq 0$ or, in other words, taking into account
the Hall effect and gradient of electron pressure showed that the process
modeled is highly dynamical and complex \citep{A-K-1} and it is virtually
impossible to distinguish between the contributions of different phenomena to
the observed scenarios. The difficulties can be overcome only by a stepwise
consideration of progressively more complex situations. Here, as in
\inlinecite{A-K-SAO}, we omit the Hall effect and electron-pressure terms in
the equations [i.e. we use the system of equations referred to as the standard
MHD model \citep[see, e.g.,][]{arber2006}].

 Thus, at least three  groups of  physical phenomena can manifest themselves in
our direct simulations: primary waves of steadying processes,
 pinch sausage instability
and turbulence of various kinds, including small-scale waves and shocks due to
the nonlinearity of the processes.

Simulations by \inlinecite{A-K-SAO} revealed a time interval during which
cross-shaped transient structures are present. Here, we use a higher-accuracy
algorithm to demonstrate that the transient structures can be accompanied by
impulsive generation of suprathermal protons in a manner similar to the
formation of proton beams in a laboratory Z-pinch.

To conclude this section, it is worth noting that our approach differs from
those typically adopted. Many researchers avoid
considering the effects of spontaneously developing small-scale turbulence,
waves and shocks because of severe difficulties entailed by such
efforts.\footnote{For example, \inlinecite{arber2006}, noting the difficulties
with the inclusion of small-scale effects in direct large-scale MHD
simulations, artificially introduce a sort of relaxation\,--\,in particular, to
suppress unresolved acoustic modes, thermal conduction and shock dissipation on
small scales. In this way, they succeed in their simulations with chromospheric
neutral particles taken into account.} In contrast, our method is specifically
tailored for describing small-scale inhomogeneities irrespective of their
scale.

\section{Mathematical Formulation of the Problem}
\label{problem}
\subsection{The System of MHD Equations}
\label{system}

To nondimensionalise the variables, we use two basic units, viz. the
number density [$N_{*}$] of either component of the quasi-neutral
proton--electron plasma and the plasma temperature [$T_{*}$], which is
characteristic of a certain layer of the solar atmosphere before the
arrival of the magnetic field and refers to the nonmagnetic
regions after the field arrival.
 Then $\rho_{*}\equiv m_\mathrm i N_{*}$ will be the unit
density. We also choose the scale height of the atmosphere
[\mbox{$H=kT_*/(m_\mathrm i g)$} (where $k$ is the Boltzmann constant
and $g$ is the gravitational acceleration)] as the unit length $L$.

Our problem consists in studying the processes that can occur in such
a layer under the influence of a magnetic field. As a unit magnetic
field, we use its characteristic magnitude [$B_0$]. Then the
combinations ${B_0}^2/(4\pi)$ and \mbox{$v_{0*}\equiv
B_{0}/\sqrt{4\pi \rho_{*}}$} can be used as the unit pressure and
velocity, respectively. We choose $t_{0*}\equiv H/v_{0*}$ as the unit
time and $cB_0/(4\pi H)$ as the unit of the electric current. Thus,
all of the units are specified, given the dimensional quantities
$T_{*}$, $N_{*}$, and $B_0$.

However, we should keep in mind that the response of the plasma
medium to the initially present magnetic field (with a
characteristic value of $B_0$) can conveniently be judged by
the value of the plasma-$\beta$ parameter,
\begin{equation}\label{beta}
\beta_{0*}=8\pi{\cal P}_*/{B_0}^2;
\end{equation}
it gives a general characterisation of the situation, being the
ratio of the characteristic dimensional pressures\,--\,the gas
pressure ${\cal P}_*=kN_{*}T_{*}$ to the magnetic pressure
$B_{0}^2/8\pi$. We write $B_0$ and, accordingly, $v_{0*}$ in
terms of $\beta_{0*}$ (recall that $T_{*}$ and $N_{*}$ do not
depend on the magnetic field) as
\begin{eqnarray}\label{edB}
B_0={\beta_{0*}}^{-1/2}\sqrt{8\pi kN_*T_*} ,\\ \label{edV}
v_{0*}={\beta_{0*}}^{-1/2}\sqrt{2kT_*/m_\mathrm i},
\end{eqnarray}
to express all the introduced units in terms of $T_{*}$,
$N_{*}$, and $\beta_{0*}$.

We restrict ourselves to a two-dimensional geometry and assume that the
physical quantities do not vary along the straight, parallel, horizontal
magnetic-field lines, and the plasma motion is directed across them. In this
case, the complete system of self-consistent nonlinear equations of collisional
magnetogasdynamics with allowances
 for finite electrical resistance, thermal conductivity,
and plasma gas-dynamical viscosity \citep{BM,Brag},
written in terms of the
dimensionless functions $\mathbfit{B}$, $T$, $\rho$, ${\cal P}$,
$\mathbfit{v}$ and the dimensionless electric current [$\mathbfit
j$], has the form
\begin{eqnarray}\label{19}
\rho\left(\frac{\partial{\mathbfit v}}{\partial t}+(\mathbfit
v\cdot\nabla)\mathbfit v\right)=-\nabla\left({\cal
P}+\frac{B^2}2\right)+M\Delta\mathbfit v,\\ \label{17}
\frac{\partial{\rho}}{\partial t}+\nabla\cdot(\rho\mathbfit
v)=0,\\  \label{sost}
 {\cal P}=\frac {\beta_{0*}}2\rho T,\\ \label{1}
\frac{\partial{\mathbfit B}}{\partial t}=\nabla\times(\mathbfit{
v\times B})-\nabla\times(\Theta\mathbfit j),\quad\mathbfit
j=\nabla\times\mathbfit B;
\\ \label{teplo}
\frac{\beta_{0*}}{2(\gamma-1)}\rho\left(\frac{\partial
T}{\partial t}+(\mathbfit v\cdot\nabla)T\right)+{\cal
P}\nabla\cdot\mathbfit v= \nonumber
\\ \nabla\cdot(K\nabla T)+\Theta j^2.
\end{eqnarray}
Here, $\gamma=5/3$; the functions
\begin{eqnarray}\label{TetK}
\Theta&=&\theta_*\beta_{0*}^{1/2}T^{-3/2},\\ \label{KTet}
K&=&\kappa_*\beta_{0*}^{3/2}T^{5/2},\\ \label{Mvis}
M&=&\mu_*\beta_{0*}^{1/2}
\end{eqnarray}
are, respectively, the local magnetic diffusivity, dimensionless local thermal
conductivity and dimensionless kinematic viscosity coefficient (assumed
to be independent of the local temperature variations); the dimensionless
factors $\theta_*$, $\kappa_*$, $\mu_*$ do not depend on $\beta_{0*}$, being
determined in Appendix A by the originally specified dimensional parameters of
the medium\,--\,specifically, $\theta_*$ depends on $T_*$, and both $\kappa_*$
and $\mu_*$ depend on $T_*$ and $N_*$. \footnote{In their Sections 2.3 and 2.4,
\inlinecite{BM} use the system of Equations~(\ref{19})\,--\,(\ref{teplo}) to investigate
 the plasma and magnetic-field dynamics without taking into
 account the Hall effect. Thus, this system describes an isotropically conducting plasma. If, however,
 this effect is included, the system should be supplemented with terms with a parameter  $\xi$ defined by Equation (\ref{xi})
 and characterising the difference between the macroscopic (dimensionless) velocities
 of the ion and electron gases, $\mathbfit v_\mathrm i-\mathbfit v_\mathrm e=\xi\mathbfit j/\rho$;
 see Sections~1.2~and~2.5 of \inlinecite{BM}. The presence of these terms implies a tensor form
 of the transport coefficients, which is equivalent to the appearance of the $\xi$-dependent
 Hall and Pedersen conductivities, $\sigma_\mathrm{H}$ and $\sigma_\mathrm{P}$. Some important  phenomena discovered
 by \inlinecite{BM} could be accounted for in terms of the local variations of $\sigma_\mathrm{H}$ and $\sigma_\mathrm{P}$
 in the course of the plasma and magnetic-field coevolution \citep{MS,APISMA}.}
We do not take into account thermoelectricity and neglect the influence of gas-dynamic
viscosity on plasma heating. As already noted, the term taking into account the
effect of gravity is omitted in Equation (\ref{19}). The equation
$\nabla\cdot\mathbfit B=0$ is automatically satisfied for the assumed
two-dimensional geometry.

We will consider here the dependence of the variables on two spatial
coordinates in a plane perpendicular to $\mathbfit B$, in which we introduce a
Cartesian coordinate system $(x,z)$ with the $x$-axis directed vertically
downward. The distribution of $B(x,z)$ over the plane is displayed as a
greyscale map; we note that, in the geometry considered, the contours of the
magnetic-field strength coincide with the electric-current lines.

\subsection{Initial Conditions}
\label{initial}

\begin{figure}
   \centerline{\hspace*{0.015\textwidth}
               \includegraphics[width=0.95\textwidth,clip=]{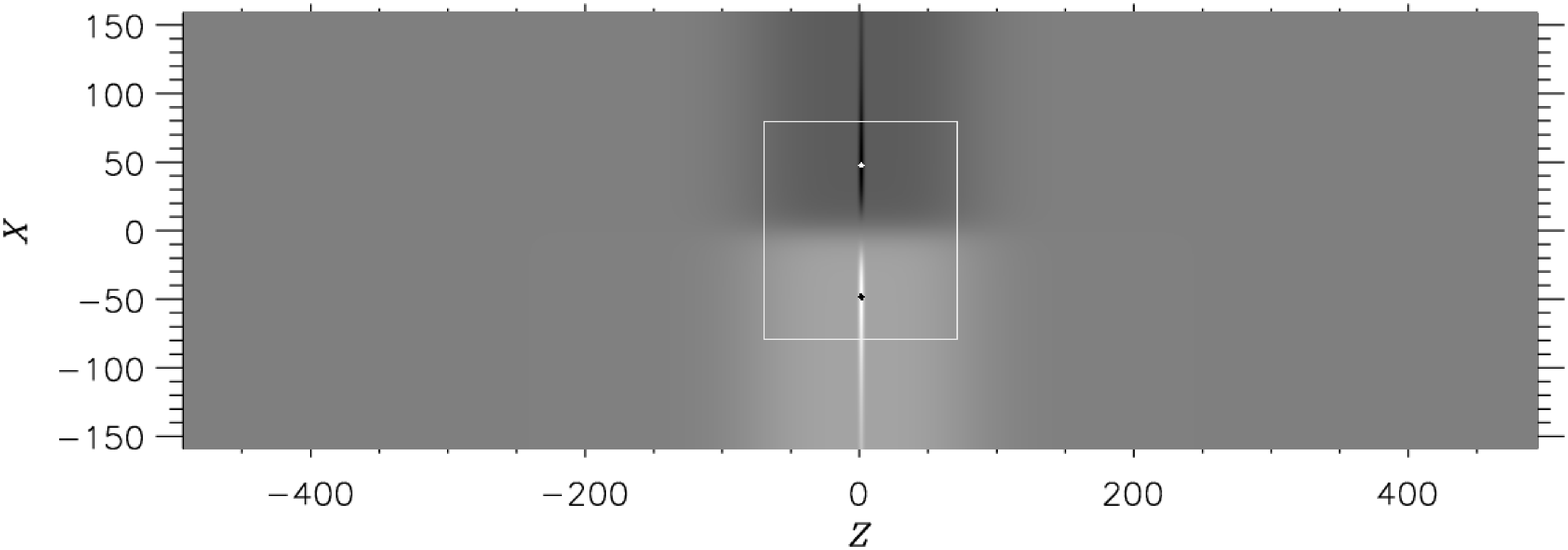}
               \hspace*{-0.03\textwidth}
              }
       \vspace{-0.25\textwidth}   % Shift close to the panel top     % EDT =edit
     \centerline{\Large \bf     % Includes the labels (here needs the color
                                %   package, see beginning of this file)
      \hspace{0.0 \textwidth}  \color{white}{(a)}
      \hspace{0.13\textwidth}  \color{white}{(a)}
         \hfill}
   \vspace{0.2\textwidth}%  %good
   \centerline{\hspace*{0.015\textwidth}
               \includegraphics[width=0.95\textwidth,clip=]{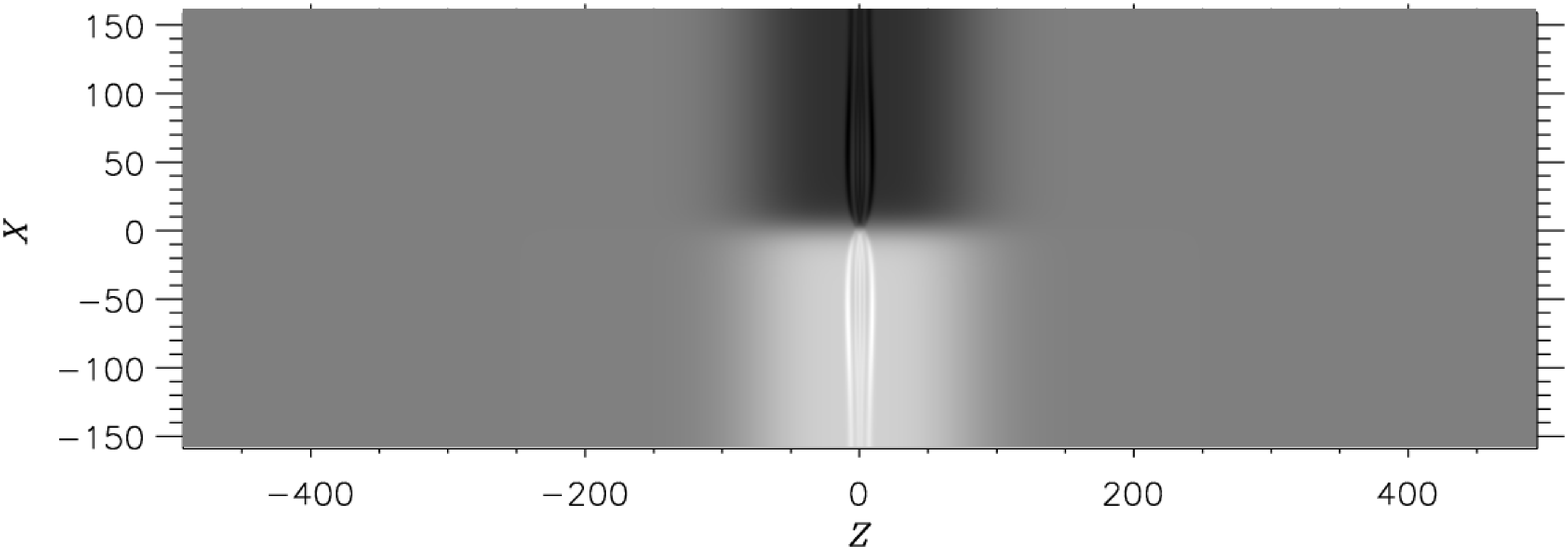}
               \hspace*{-0.03\textwidth}
              }
     \vspace{-0.25\textwidth}   %
     \centerline{\Large \bf     % Includes the labels (here needs the color package)
      \hspace{0.0 \textwidth} \color{white}{(c)}
       \hspace{0.13\textwidth}  \color{white}{(b)}
         \hfill}
    \vspace{0.2\textwidth}%
 \centerline{\hspace*{0.015\textwidth}
               \includegraphics[width=0.95\textwidth,clip=]{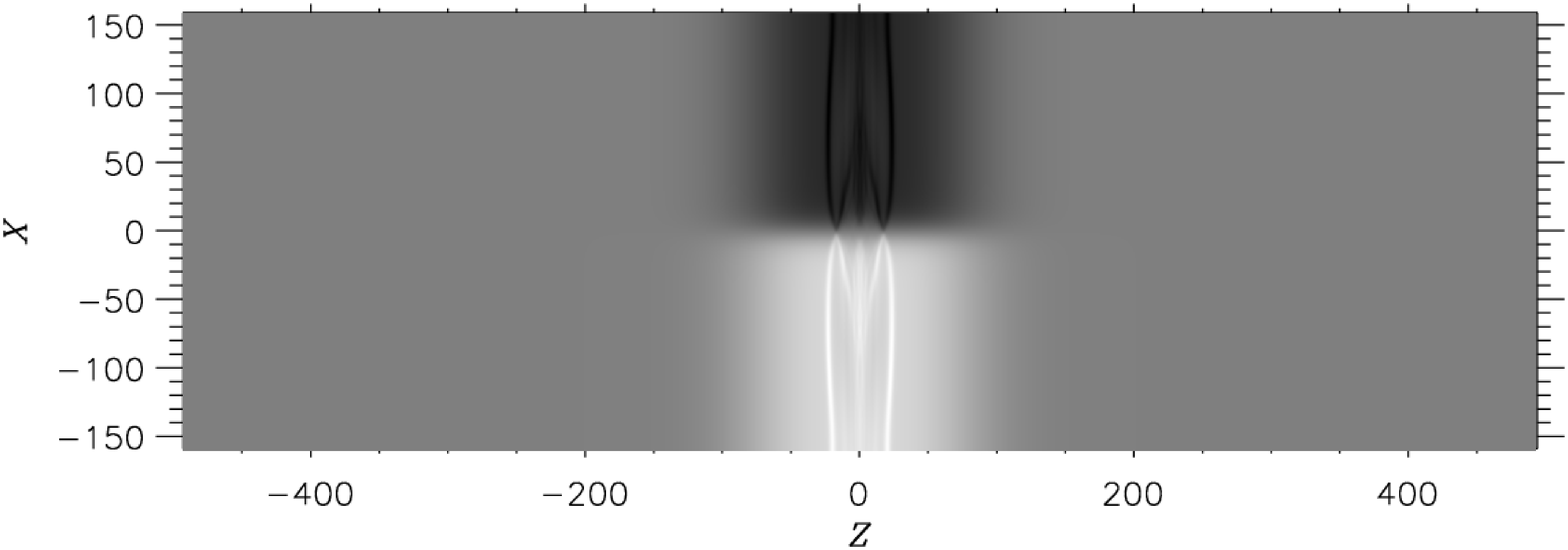}
               \hspace*{-0.03\textwidth}
              }
    \vspace{-0.25\textwidth}
     \centerline{\Large \bf     % Includes the labels (here needs the color
                                %   package, see beginning of this file)
      \hspace{0.0 \textwidth}  \color{white}{(a)}
     \hspace{0.13\textwidth}  \color{white}{(c)}
         \hfill}
     \vspace{0.21\textwidth}
\caption{Distribution of $B(x,z)$ in the plane $(x,z)$ for the outset stage
of the magnetic field -- plasma coevolution at $\beta_{0*}=1.5$ in the
absence of gas-dynamic viscosity at different times: (a) $t=0$, the  cases of both PIB
 and NIB (see below in this section); (b) $t=0.008$ and (c) $t=0.024$, the case of PIB. The
 $x$-axis is directed downward from the upper left corner, and the
 $z$-axis is directed to the right. The quantities $X$ and $Z$ are related to $x$ and $z$
  according to Equations (\ref{xxczzc}).
 The image is stretched
 vertically by a factor of 4.4. Light: positive magnetic field
 (directed inward); dark: negative magnetic field.
 The white dot on the dark background and the black dot on the
 light background mark the positions of the minimum and maximum
 $B(x,z)|_{t=0}$, respectively. [The other greyscale maps represent the central
 part of this domain, marked with the white frame in panel (a).]}\label{f1}
\end{figure}

Let the absolute value of the initial magnetic  field be symmetric with respect
to two symmetry axes, $x=x_\mathrm{c}=x_\mathrm{max}/2$ [where $B(x_\mathrm
c,z)|_{t=0}=0$] and $z=z_\mathrm{c}=z_\mathrm{max}/2$. The distribution of
$B(x,z)|_{t=0}$ is shown in Figure \ref{f1}a, where the following notation is
used:
\begin{equation}\label{xxczzc}
X=(x-x_\mathrm{c})/x_\mathrm{max},\ Z=(z-z_\mathrm{c})/z_\mathrm{max}.
\end{equation}
 An analytic formula that
specifies the initial magnetic field [$B(x,z)|_{t=0}$] shown in Figure
\ref{f1}a is given in Appendix~B.
The  initial field thus chosen is the sum of two terms (see Appendix~B): the
first term is smaller in absolute magnitude and is distributed over the
light grey and dark grey areas in Figure~\ref{f1}a; the second one is greater and
more localised (narrow black and white strips in the figure). Since the
equation \mbox{$\nabla\cdot\mathbfit B=0$} is automatically satisfied in the
two-dimensional geometry of the problem, each of these terms can be considered
a separate magnetic field, either a distributed or a more localised one.

We
assume that the plasma is initially motionless and isothermal:
\begin{equation}\label{VTnach}
\mathbfit v(x,z)|_{t=0}=0,\quad T(x,z)|_{t=0}=1.
\end{equation}
In this case, according to Equation (\ref{sost}), the density at $t=0$ differs
from the pressure only by a constant factor. The initial condition
for the pressure should specify, however, the degree of balance
between the initial distributions of the magnetic and gas pressure.
Here, as in \inlinecite{A-K-SAO}, we restrict ourselves to two cases.

\textit{The case of no initial balance (NIB)}. If the pressure
of the initially specified magnetic field $B(x,z)|_{t=0}$ is
not balanced by the gas pressure, then, according to our choice
of units,
\begin{equation}\label{Rho1}
\rho(x,z)|_{t=0}=1
\end{equation}
and, accordingly, in view of the Equation (\ref{sost}) of state,
\begin{equation}\label{P1}
 {\cal P}(x,z)|_{t=0}=\beta_{0*}/2.
\end{equation}

\textit{The case of partial initial balance (PIB)}.  We assume
that, in the case of PIB, the distributed magnetic field, being
in balance with the local gas pressure ${\cal P}(x,z)|_{t=0}$,
is superposed with an unbalanced localised magnetic field and
the balance thus appears to be only partial. Physically, this
case can be interpreted as the penetration of a new, unbalanced
magnetic field to the area of the old, pre-existing field
(which has already reached a pressure balance). The
corresponding formulas for the initial pressure and density are
given in Appendix~C.

\section{The Properties of the Exact Solution Dictated by
the Formulation of our Initial-Value Problem.} \label{analyt}

Our analyses of the form of Equations (\ref{19})\,--\,(\ref{teplo}) with
initial conditions given by Equations (\ref{VTnach}), (\ref{nachvid}) together
with either Equations (\ref{Rho1}), (\ref{P1}) or (\ref{P2}), (\ref{Rho2}),
 lead to the following conclusions.

i) If the initial magnetic and gas pressures are completely balanced,  a
stationary solution exists. However, in both the cases of a partial balance
(PIB) and of no initial balance (NIB), dynamical processes manifest themselves.
We discuss them now assuming that $M=0$ in Equation (\ref{19}).

ii) If the initial configuration of the magnetic field has the form
 of Equation~(\ref{nachvid}), illustrated in Figure~\ref{f1}a, the subsequent evolution of the
magnetic field and plasma state should be similar to those in a laboratory
Z-pinch. Note, however, that the contraction of the electric current to the
axis $x=x_\mathrm c$ does not result here in substantial increases in the
maximum $|\mathbfit B|$ in view of the planar geometry.

iii) In the NIB case, the coevolution of the plasma and $\mathbfit
B$ begins with the origin of a system of powerful waves, fast
electric-current contraction, collapse of the plasma and its
subsequent retreat. All of these processes strongly affect the
bulk of plasma.

iv) In the PIB case, the early evolution is more quiet.
 The unbalanced
magnetic-field component $d$ present between the layers of the balanced field
(Figure~\ref{f1}a) form ``tongues'' stretching to $x=x_\mathrm c$, which
provoke the sausage instability. Since $\mathbfit B=0$ at $x=x_\mathrm c$,
an $x$ component of the magnetic-pressure gradient is present near this axis;
according to Equation (\ref{19}), oppositely directed flows with a velocity $v_x$,
directed toward this axis from the both sides, develop. The form of
Equation~(\ref{1}) indicates that the magnetic field is partially frozen in the
plasma. Therefore, these flows carry the magnetic field closer to the axis,
which results in a further steepening of the magnetic-field gradients, and so they intensify
themselves. The substance can freely flow out along the neutral line and then
spread from it (no electrodes or walls are present, in contrast to the
conditions of laboratory setups). As to magnetic viscosity [$\Theta$], it is not
effective damper of the process, since the great gradients of $B$ correspond to
great $j$, i.e. enhanced heating of plasma and decreasing of $\Theta$ in
accordance with Equations (\ref{teplo}, \ref{KTet}). Therefore, in the absence of
gas viscosity $M$, the initial-value problem of Equations (\ref{19})\,--\,(\ref{teplo}),
(\ref{VTnach}), (\ref{nachvid}), (\ref{Rho1}), (\ref{P1}), or (\ref{P2},
\ref{Rho2})
 has solutions with the velocity [$v$] approaching $\infty$
with $t$. The velocity increases the faster, the closer the
magnetic``tongues'' to the axis $x=x_\mathrm c$. It is difficult to
predict whether this process will be monotonic or impulsive.

v) The most important feature of  the  process considered is
the presence of opposite plasma flows [$v_x$]
approaching each other through the zone where $B\approx 0$.
 It is therefore reasonable to conjecture that the
inclusion of the gas viscosity [$M$] in Equation~(\ref{19}) can slow down or
even stop the process. (In contrast to the magnetic
viscosity [$\Theta$], the gas viscosity increases with temperature
according to Equation (\ref{mun})). Here, we are interested in a qualitative
evaluation of the role of the gas viscosity and ignore, for
simplicity, its local variations with temperature.

 \section{The Procedure of Numerical Solution}
\label{numer}

The verification criterion for a numerical code
solving the two-dimensional initial-value problem in Equations
(\ref{19})\,--\,(\ref{teplo}), (\ref{VTnach}), (\ref{nachvid}),
(\ref{Rho1}), (\ref{P1}), or (\ref{P2}, \ref{Rho2}) is the similarity
between the numerical and exact  solutions.
However, the numerical solution possesses some features that
are not present in the exact solution. For
example, the exact solution of the problem in Equations
(\ref{19})\,--\,(\ref{teplo}), (\ref{VTnach}), (\ref{nachvid}),
(\ref{Rho1}), (\ref{P1}), or (\ref{P2}, \ref{Rho2})
 must be of the type of
symmetry relative to the axis $x=x_\mathrm c$ considered and cannot
describe the wriggle (or kink) instability of the Z-pinch,
whereas its numerical solution can; the reason is that
numerical noise prevents accurately describing the symmetry
imposed by the initial conditions.

To solve our initial-value problem numerically, we use a finite-difference
scheme that approximates the equations with the second order of accuracy in
time and space; central differences are used to approximate the spatial
derivatives. It is known that stable numerical methods of second-order accuracy
introduce no spurious dissipation \citep[although they may introduce spurious
dispersion; methods of a first-order accuracy introduce spurious dissipation
without spurious dispersion\, --\, see][]{LeVeque}; on this basis, many
numerical algorithms free of spurious dissipation have been developed
\citep{ThomasR}. Our second-order-accuracy technique belongs to this class of
methods free of numerical dissipation.

Our scheme is constructed so as to ensure the existence of a
non-increasing grid analogue of the wave-energy functional; this
immediately proves a stability theorem for the method. The
scheme is conservative, i.e. the energy, mass and momentum are
conserved. As already noted, the solution of the problem can be
sought in the class of generalised functions; the algorithm is
able to automatically make transitions from smooth to
generalised solutions and vice versa. Therefore, the method
makes it possible to obtain non-differentiable solutions, in
particular, with discontinuities (even multiple). This property
of our numerical method proves to be important, since we will
study the breakdown of the initially smooth field into
small-scale waves and other structures, i.e. into the features
that can be described either by functions with large
derivatives or even by non-smooth functions. Our numerical
method generalises the previously developed method for the
numerical integration of equations of motion of the atmospheric
gas \citep{Ksh-Gavr}. Structurally, the numerical-integration
formulas resemble those of the Lax--Wendroff method. A
comprehensive analysis of our numerical technique was given by
\cite{Ksh06}.

 Our simulations are carried out  under the assumption that the
velocity component normal to the boundary of the computation
domain and the normal derivative of the magnetic field vanish
at this boundary. The computation domain is chosen to be large
enough to eliminate boundary effects affecting the processes in
its central part, where the phenomena under study develop.

We solve the problem for a chromospheric layer where the temperature
of both the ion (proton) and electron components of the plasma is
25,000~K, which gives a total characteristic temperature of
$T_*=50,000$~K for the plasma medium. We assume here
\mbox{$N_*=10^{15}\ \mathrm m^{-3}$} \citep{DEM}. In such a layer,
$\theta_{*}=1.2\times 10^{-8}$, $\kappa_{*}=3.8\times10^{-4}$, and
$\mu_*=5\times 10^{-6}$ (see Appendix~A).

Further, we consider the case of $\beta_{0*}=1.6$ turning sometimes to those of
$\beta_{0*}=1.5$ or $\beta_{0*}=2.3$. In accordance with Section
\ref{problem}, the dimensional units for the case of $\beta_{0*}=1.6$ are:
$B_0=0.33\ \mathrm G$,
\begin{equation}\label{Edin}
 v_{0*}=23\ \mathrm{km\,s}^{-1},\quad t_{0*}=1.1\ \mathrm{minutes}.
\end{equation}

The applicability of the our initial-value problem to description
of plasma dynamics at this intensity of  $ B_0$
is discussed in Appendix~D.

The calculations for $M=0$ were carried out in a domain with
$x_\mathrm{max}=0.207$, $z_\mathrm{max}=2.8$, which corresponds to 300~km  (320
grid points) in height and 4200~km (988 grid points) in the horizontal
direction. For $M\neq 0$, the domain measures $0.207 \times 0.639$, or $300
\times 922~\mathrm{{km}^2}$, with $80 \times 247$ grid points.

\section{Results}
\subsection{The Phenomenon of Velocity Burst}

Our simulations reveal an ensemble of primary waves produced by contraction
and expansion of the whole current zone associated with the unbalanced
component of $\mathbfit B$ (see Section~\ref{our2} and Figures~\ref{f1}b,c). Against
the background of these waves, at the centre of the magnetic configuration, the
pinch-sausage instability begins developing and becomes especially pronounced
after the waves leave this area (see Section~\ref{discuss}).

\begin{figure}
   \centerline{\hspace*{0.015\textwidth}
               \includegraphics[width=0.75\textwidth,clip=]{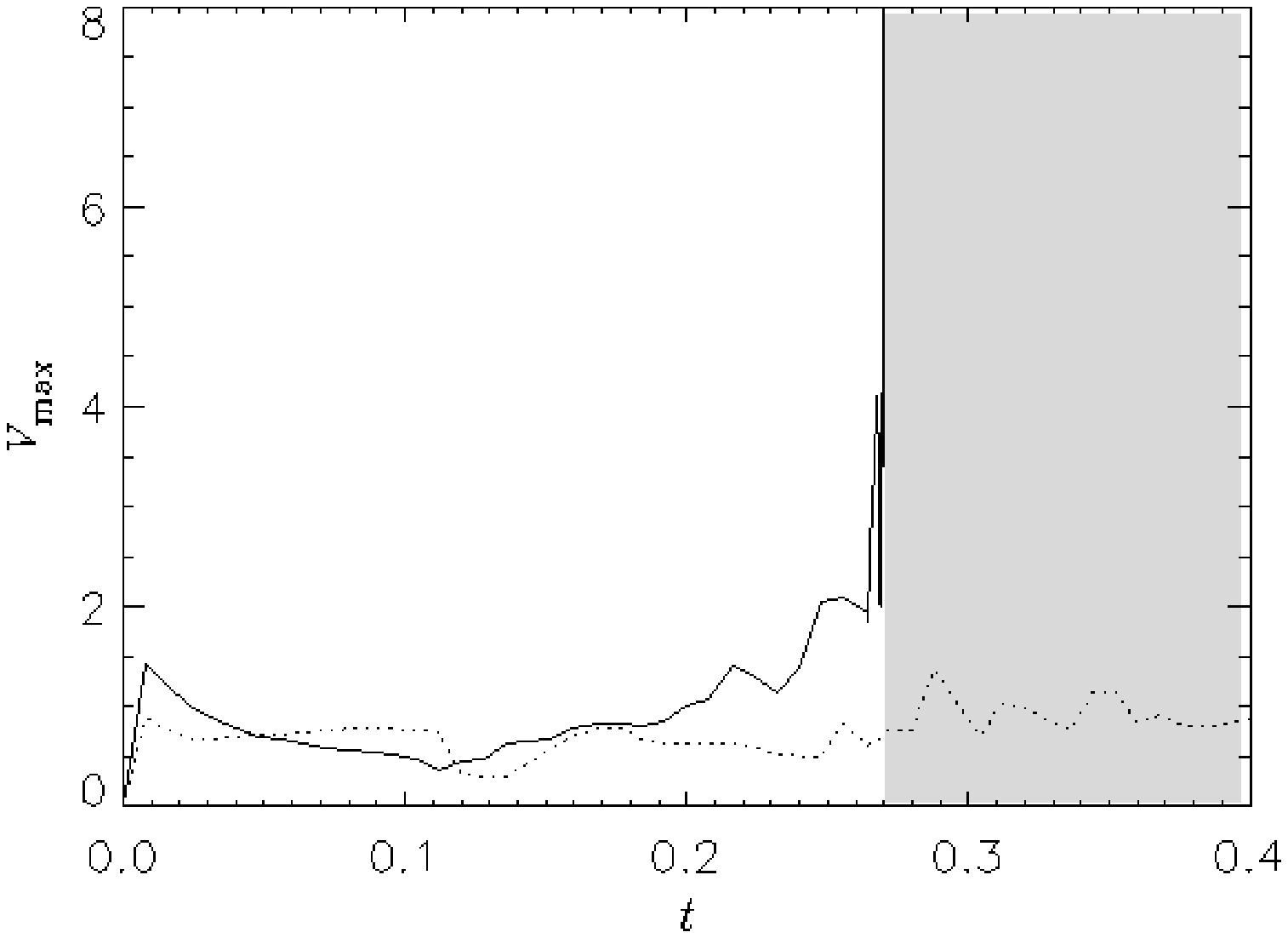}
               \hspace*{-0.03\textwidth}
              }
    \vspace{-0.40\textwidth}   % Shift close to the panel top %     PAIR WITH LLLL
     \centerline{\Large     % Includes the labels (here needs the color  package, see beginning of this file)
      \hspace{0.0 \textwidth}  \color{white}{(a)}
       \hspace{0.23\textwidth}  \color{black}{(a)}
         \hfill}
    \vspace{0.34\textwidth}    % Shift back to the panel bottom  ; PAIR WITH LLLL
   \centerline{\hspace*{0.015\textwidth}
               \includegraphics[width=0.75\textwidth,clip=]{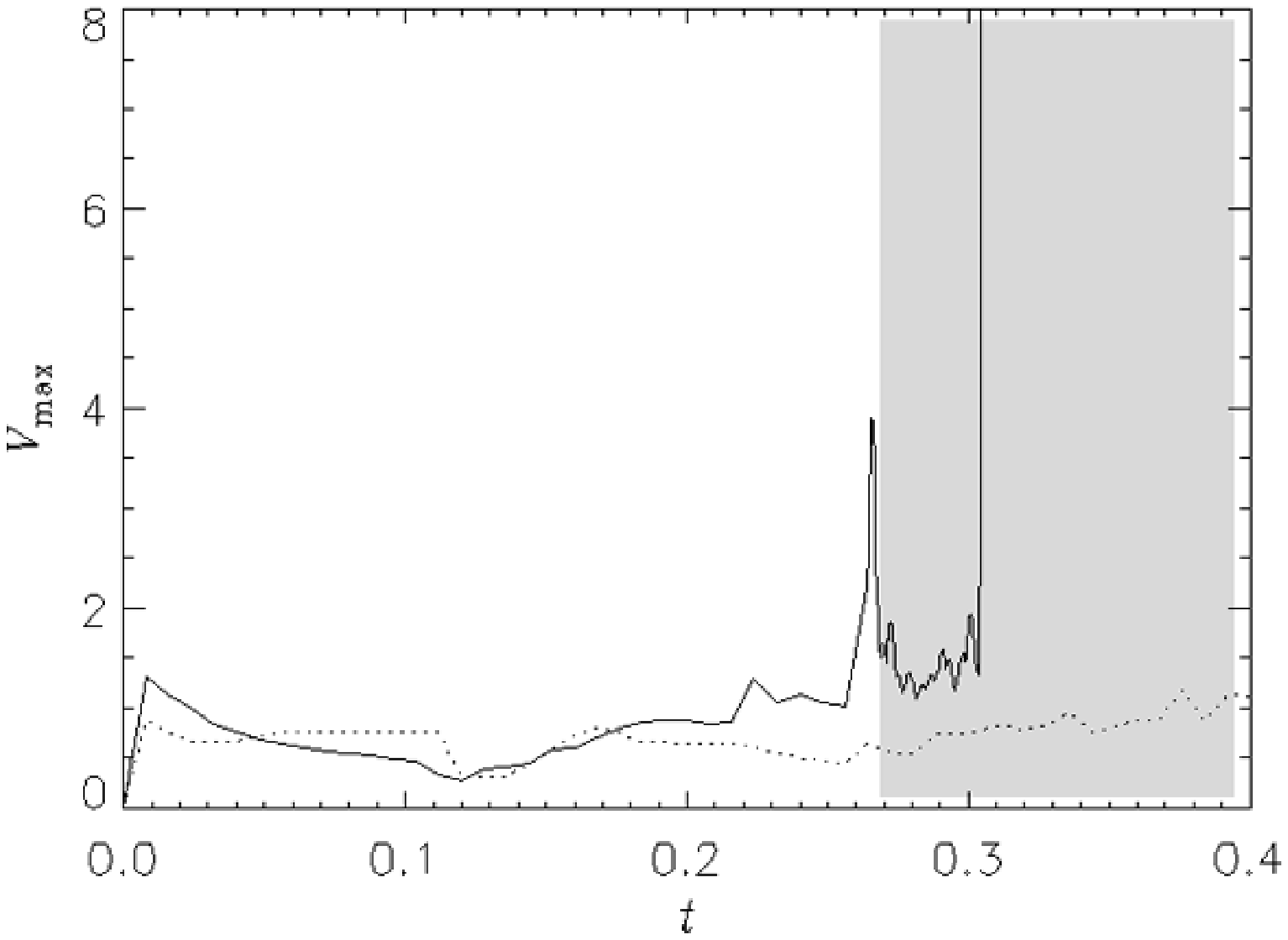}
               \hspace*{-0.03\textwidth}
              }
     \vspace{-0.40\textwidth}
     \centerline{\Large     % Includes the labels (here needs the color package)
      \hspace{0.0 \textwidth} \color{white}{(c)}
       \hspace{0.23\textwidth}  \color{black}{(b)}
        \hfill}
       \vspace{0.296\textwidth}
 \centerline{\hspace*{0.038\textwidth}
               \includegraphics[width=1.02\textwidth,clip=]{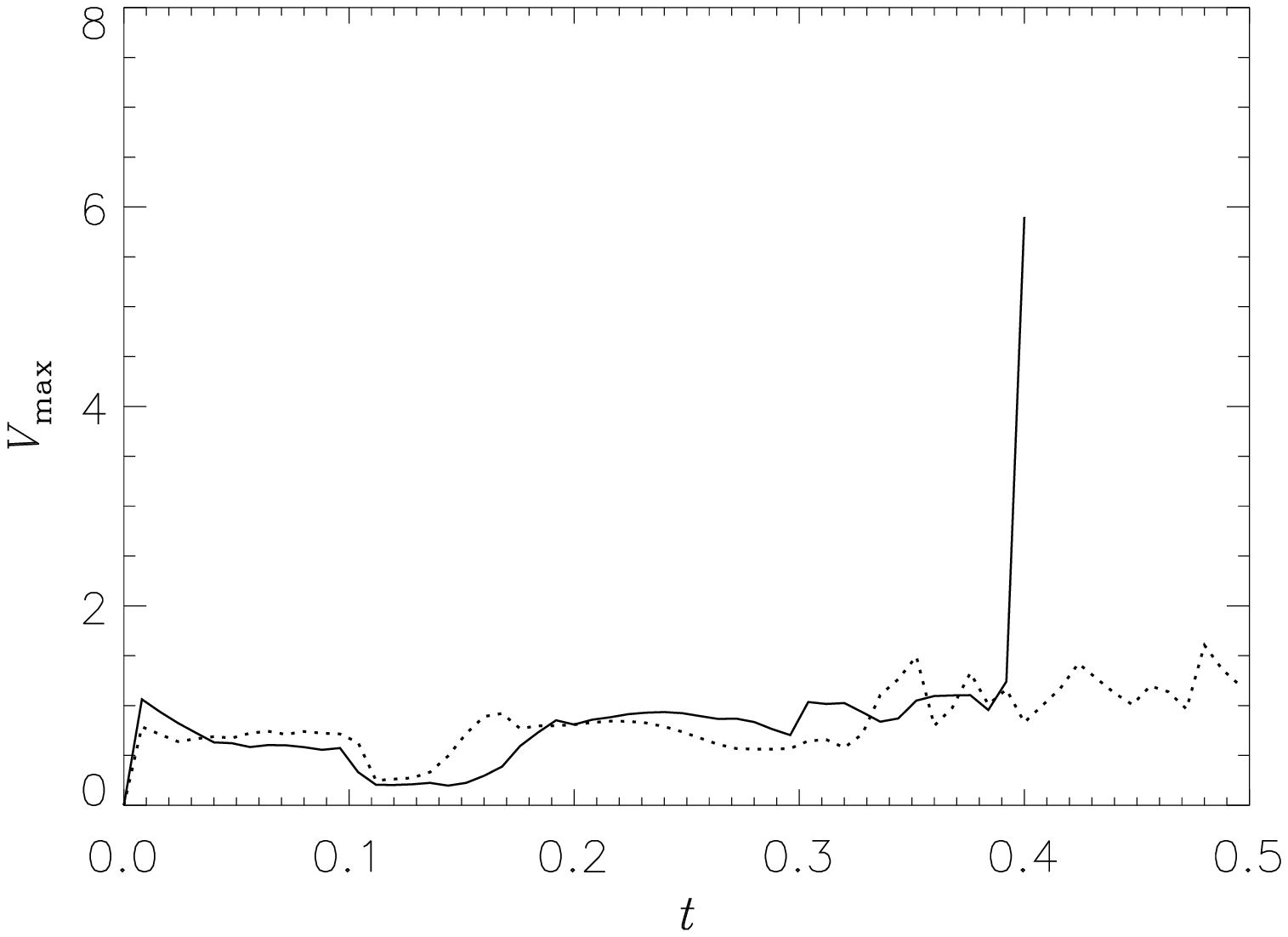}
                  \hspace*{-0.072\textwidth}
              }
    \vspace{-0.49\textwidth}
     \centerline{\Large     % Includes the labels (here needs the color %   package, see beginning of this file)
      \hspace{0.0 \textwidth}  \color{white}{(a)}
     \hspace{0.235\textwidth}  \color{black}{(c)}  % more value - more right
         \hfill}
    \vspace{0.43\textwidth}
\caption{The maximum absolute velocity over the computation
domain as a function of time in the cases of PIB (solid curve) and
NIB (dotted curve) at $\beta_{0*}=1.5$ (a), $\beta_{0*}=1.6$ (b)
and $\beta_{0*}=2.3$ (c) in the
absence of gas viscosity [$M$].}\label{f2}
\end{figure}

As shown previously \citep{A-K-SAO}, the nonlinear stage of the pinch-sausage
instability forms cross-shaped transient structures, which are more complex and
smaller in size in the calmer PIB case, as compared with the more dynamical,
completely unbalanced NIB case. The development of the pinch-sausage
instability in the PIB case passes through two early stages: only after this
pinch effect has manifested itself in the unbalanced component of the magnetic
field and has produced its cross-shaped pattern, does the  sausage instability set
in in the balanced component of the field, and the concentration of the whole
magnetic field approaches the symmetry axis $x=x_\mathrm{c}$ of the initial
magnetic configuration. This gives rise to intricate transient flow structures,
which facilitate the development of turbulence and, accordingly, favour the
heating of the plasma. Thus, $T$ can be observed to double in the PIB case,
while no appreciable temperature changes are associated with the pinch
instability in the NIB case.

Now, simulations based on the improved algorithm ensuring higher
accuracy reveal another peculiarity of the coevolution of the
magnetic field and plasma state in the PIB case (Figure~\ref{f2}). Against the
background of the above-mentioned fairly gradual variations in the
physical quantities, impulsive phenomena of increasing plasma
velocity, i.e. velocity burst, develop, no similar phenomena being
observed in the case of NIB.

The numerical result obtained here agrees with the properties of the exact
solution deducible from the mathematical form of our initial-value problem (see
Section \ref{analyt}). Thus, in the PIB case, we observe the whole scenario of
the sausage pinch instability, up to the manifestation of the final velocity
burst. (We note that no exceptions occur in the process of computation, and the
code continues running with progressively decreasing steps).

Generally, $\beta_{0*}$ is higher, the slower the process (Figure~\ref{f2}); we also observe
this regularity in other numerical simulations, not described here,
 although there is no smooth dependence of the parameter values
$\beta_{0*}$ (see Figures~\ref{f2}a,b for an example).

Physical reasons for this effect can naturally be expected, since we solve an
initial-value MHD problem with an unstable initial magnetic configuration,
where the sausage instability develops under different physical conditions due
to the presence of interfering primary waves, even at very close values of
$\beta_{0*}$ (see Section~\ref{interfer} below).

Also, no smooth transition between the final stage of the numerical solutions
 should be expected since the required
accuracy can more hardly be achieved. For example, in both cases $\beta_{0*} = 1.5$
and $1.6$, our numerical solutions represent the evolution of the process with
the gradual velocity variations changing into a dramatic increase (the white
areas in Figures~\ref{f2}a,b). This increase should result in a rapid growth of
both physical and numerical instabilities with the spatial scale of
disturbances approaching zero. At this stage (the shaded areas in
Figures~\ref{f2}a,b right of about $t = 0.27$), the simulation requires
progressively shorter time steps and catastrophically loses its accuracy. Thus,
the white and shaded areas correspond to a normal and a poor accuracy,
respectively (this is why the small change in $\beta_{0*}$ can delay the final
infinite velocity growth). However, we do show the curves in the shadowed areas
in Figures~\ref{f2}a,b, since they demonstrate that our numerical solution
actually reproduces the analytically predicted velocity divergence. In no way does it give
 a quantitative description of the final stage of the processes.

\subsection{Velocity Burst and the Known Pinch-Effect Features}
\begin{figure}    %%%%%%%%%%%%%%%%%% FIGURE 3   rotation + cut
   \centerline{\includegraphics[width=0.45\textwidth,clip=]{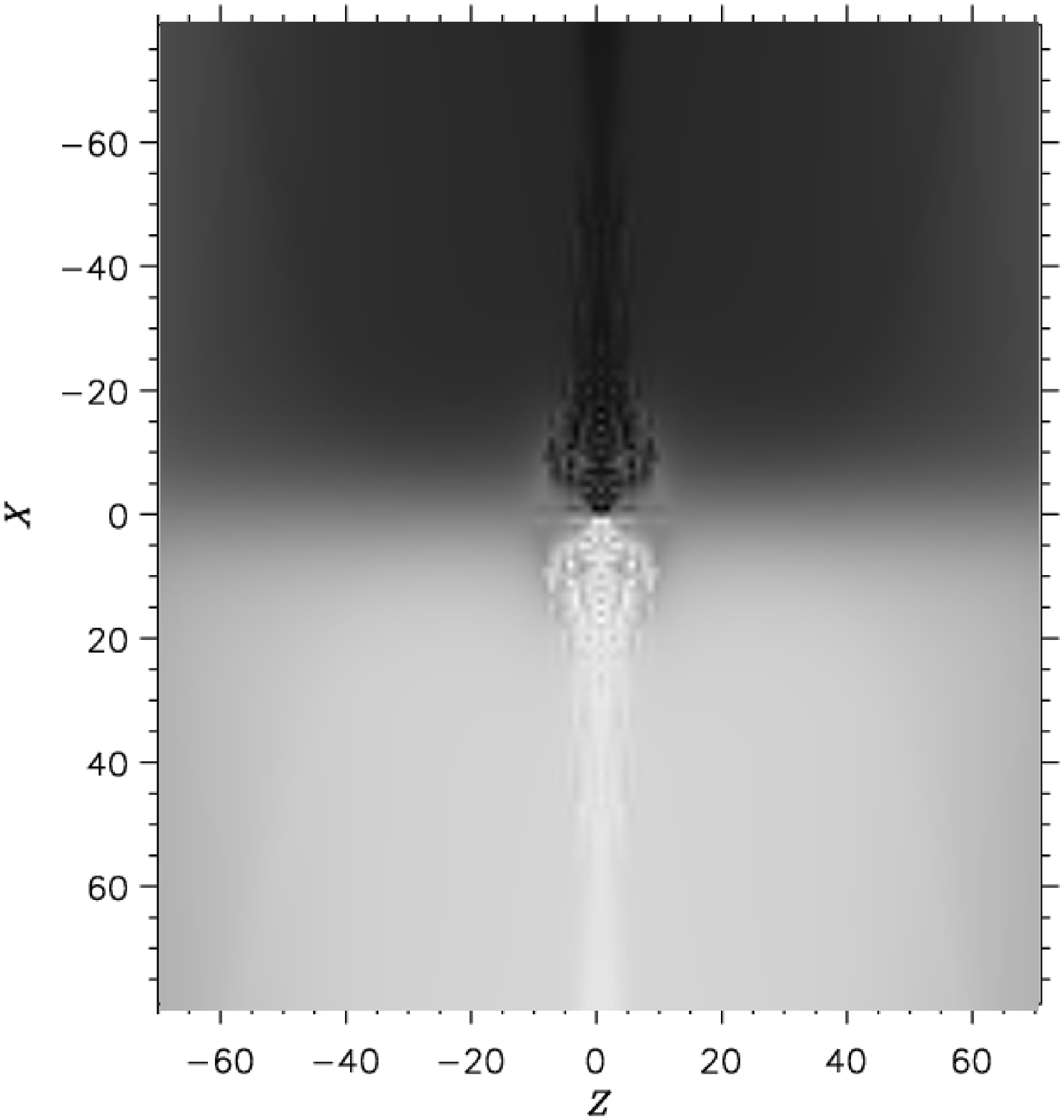}
               \includegraphics[width=0.45\textwidth,clip=]{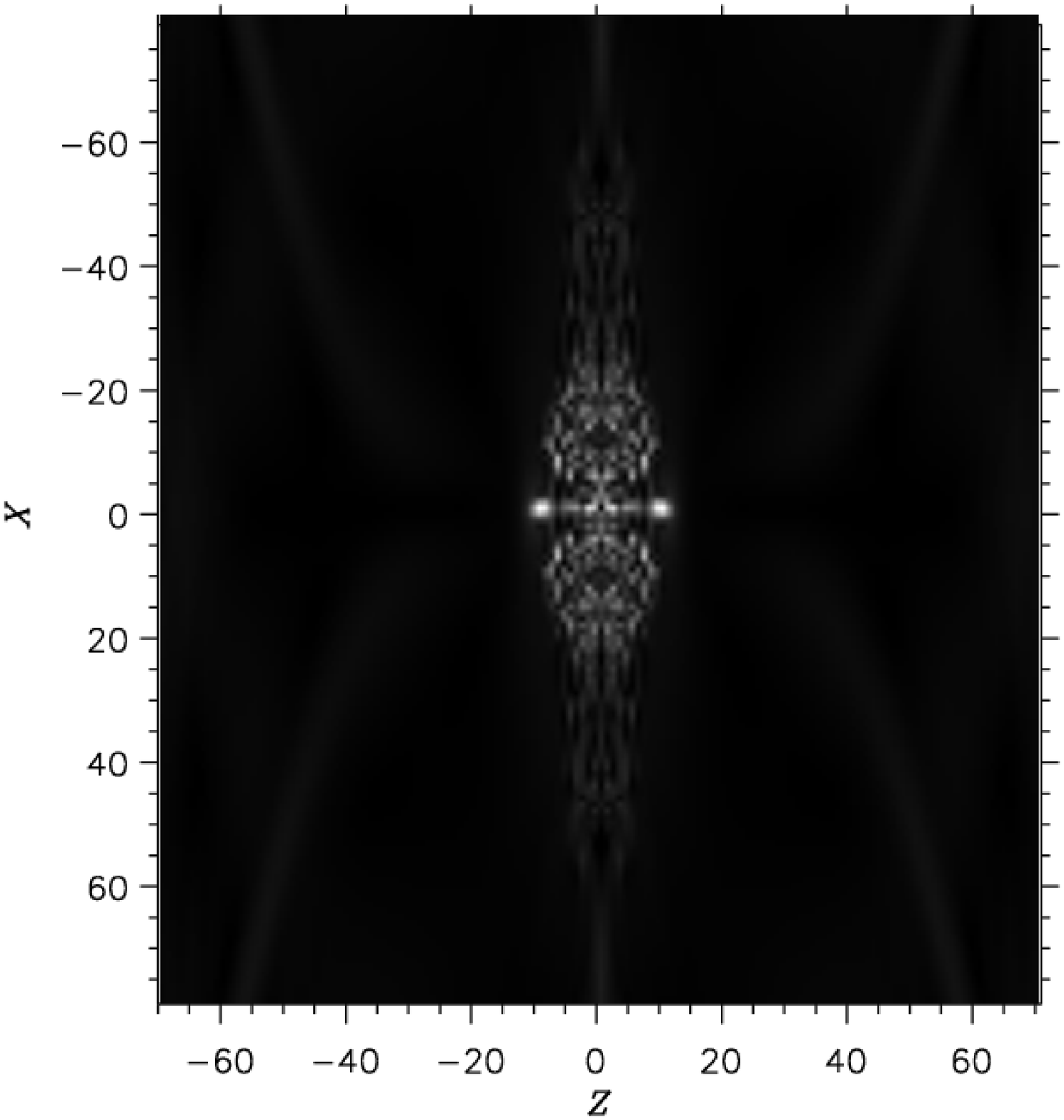}
              }
     \caption{Distribution of $B$ (left) and $v$ (right) in the plane $(x,z)$
at $t=0.216$, $M=0$.        }
   \label{f3}
   \end{figure}

 We consider the process in more detail precisely for the case  $\beta_{0*}=1.6$, where individual
peaks can be singled out. Each of them has features of an underdeveloped final
velocity burst.
The solution at $\beta_{0*}=1.6$ shows that a first, faint
velocity burst with a peak velocity of $v_1=1.2$ at $t=0.224$
precedes the second one with $v_2=3.9$ at $t=0.266$
(Figure~\ref{f2}). The first burst appears when the cross-shaped
pattern has already formed (Figure~\ref{f3}).

\begin{figure}    %%%%%%%%%%%%%%%%%% FIGURE 3   rotation + cut
% Original BoundingBox (in the .eps files) : 54 360 558 720
%   corresponds to the coordinates of: left, bottom, right, top   (unrotated)
   \centerline{\includegraphics[width=0.45\textwidth,clip=]{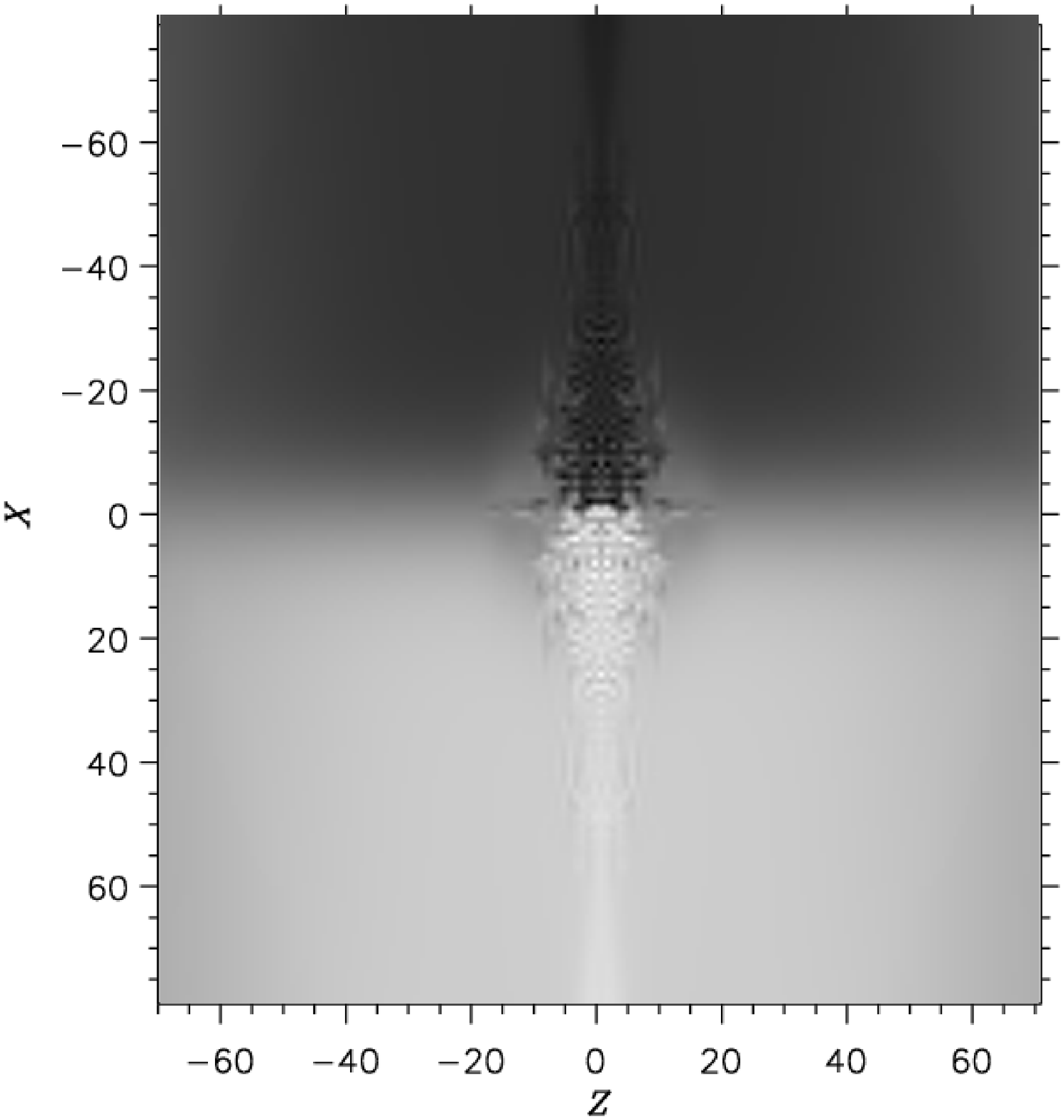}
               \includegraphics[width=0.45\textwidth,clip=]{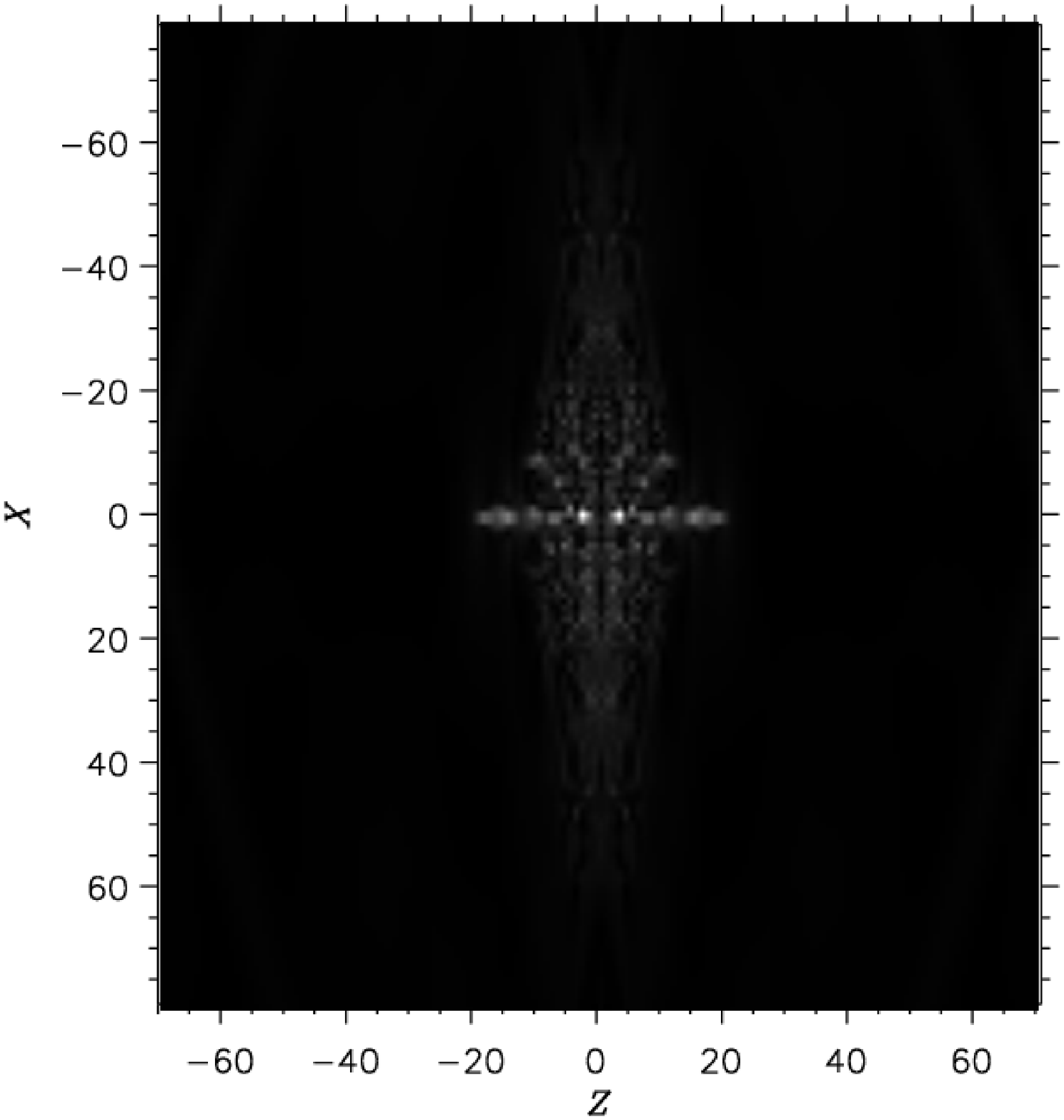}
              }
     \caption{Distribution of $B$ at $t=0.256$ (left)
and $v$ at the onset of the velocity burst, $t=0.264$,
(right) in the plane $(x,z)$.
        }
   \label{f4}
   \end{figure}

\begin{figure}    %%%%%%%%%%%%%%%%%% FIGURE 3   rotation + cut
  \centerline{\includegraphics[width=0.4\textwidth,clip=]{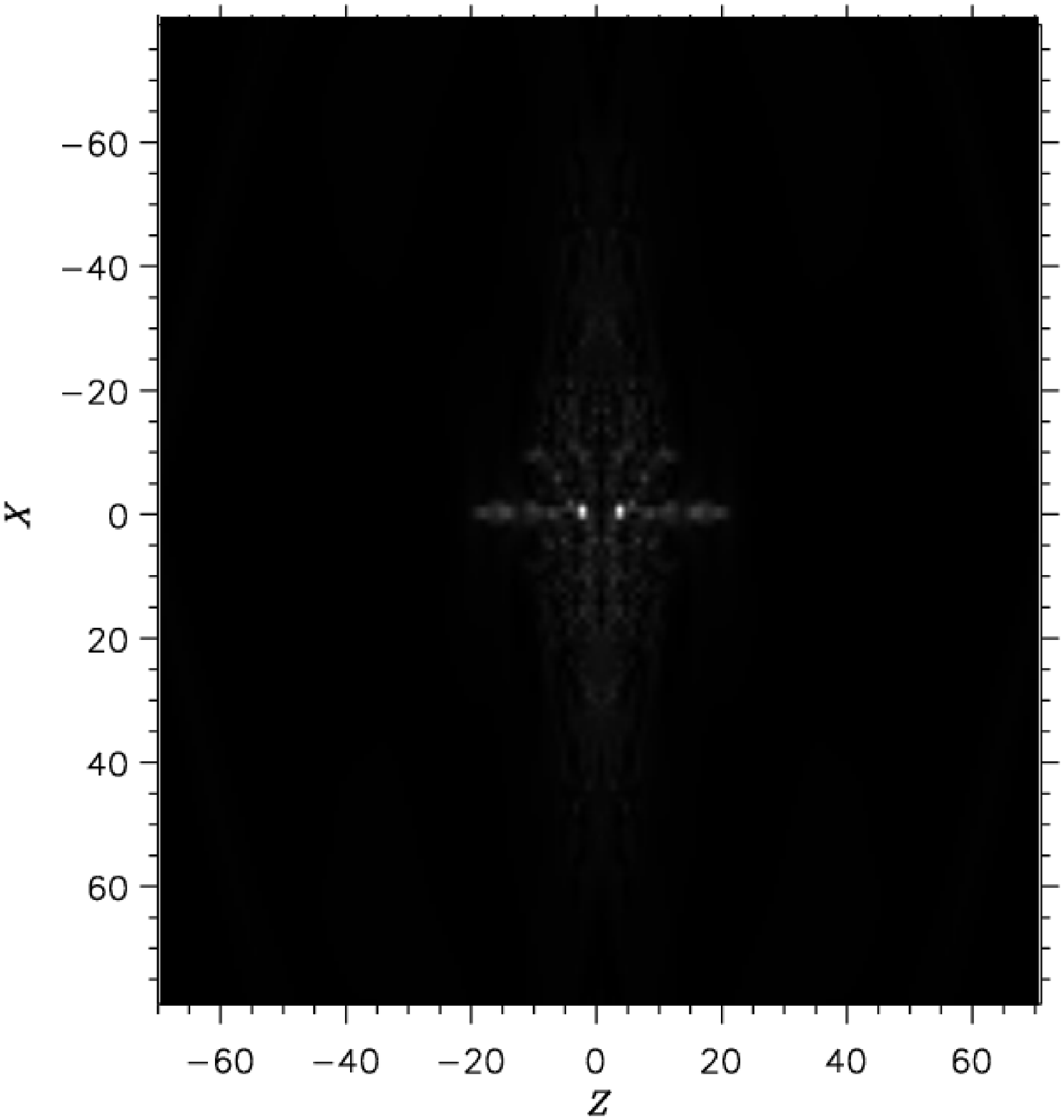}
               \includegraphics[width=0.6\textwidth,clip=]{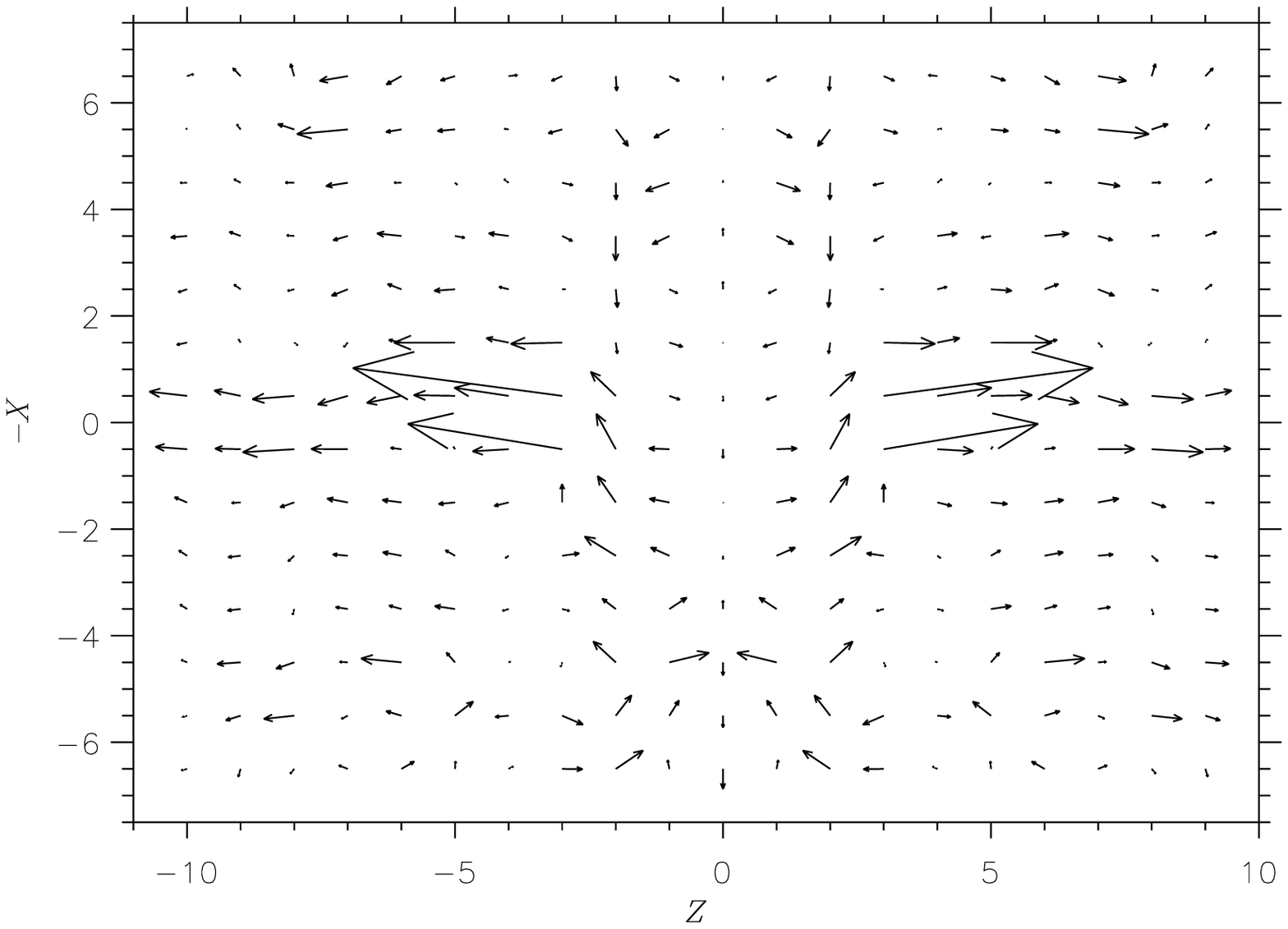}
              }
     \caption{Distribution of  $v$ (left) in the plane $(x,z)$
at the time of the second velocity peak, $t=0.2656$.
 Right:  vector map of $\mathbfit v$ (magnified) for the same time.
A division of the vertical ($x$) axis is 0.93~km;
of the horizontal ($z$) axis, 4.25~km.
The longest arrow corresponds to a velocity of $v_2=3.9$.
        }
   \label{f5}
   \end{figure}

Later, the distributed magnetic field approaches the symmetry axis
$x=x_\mathrm{c}$ to occupy an area initially free of magnetic field
(Figure~\ref{f4}, left). This induces a sharp increase in the
current-aligned velocity (which is seen as a brightening along the
axis $x=x_\mathrm{c}$ in Figure~\ref{f4}, right). The high-speed jets
corresponding to the peak of the velocity burst are located in
the central part of the magnetic configuration (Figure~\ref{f5}). This
velocity burst occurs in a time interval of duration 0.0028.

Since the  plasma  velocity $\mathbfit v$  in the  system of MHD
equations \citep{BM} coincides with the bulk velocity of the ion
component $\mathbfit v_\mathrm i$, at the velocity burst, an ion
acquires an additional kinetic energy of $m_{\mathrm i} v^2/2$. This
corresponds to the ion kinetic temperature $T_{\mathrm
{ki}}=511,000$~K gained rapidly, in 0.1~seconds. Thus, we see that the
velocity burst produces suprathermal protons.
\begin{figure}
   \centerline{\hspace*{0.015\textwidth}
               \includegraphics[width=0.75\textwidth,bb=72 0 504 360,clip=]{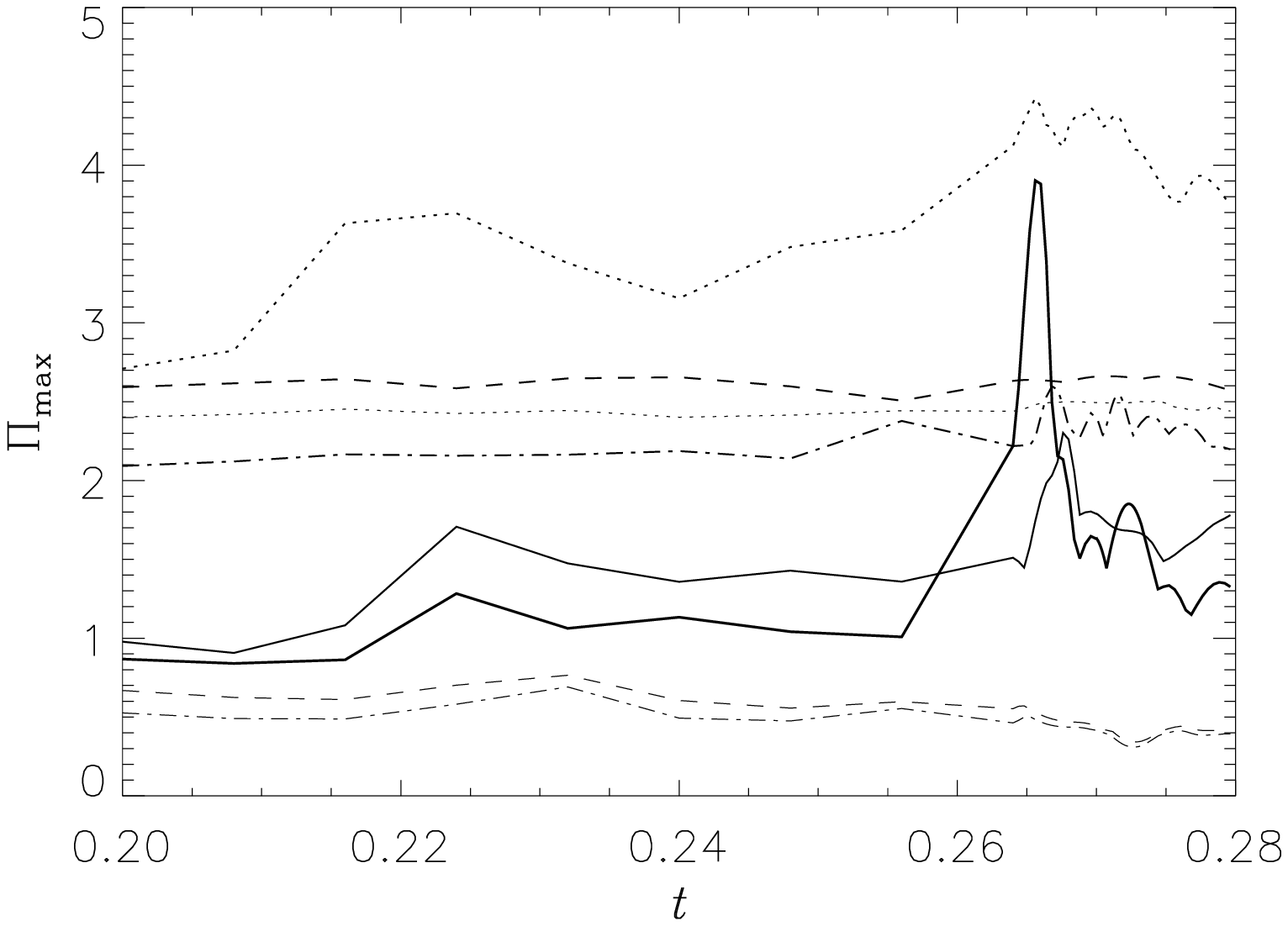}
               }
 \caption{Temporal variation of MHD quantities near the time of velocity burst:
$v_\mathrm{max}$ (heavy solid curve), $(B^2/2)_\mathrm{max}$ (light solid curve),
$2.5T_\mathrm{max}$ (heavy dotted curve), $2.5T_\mathrm{min}$ (light dotted curve),
$2.5\rho_\mathrm{max}$ (heavy dashed curve), $2.5\rho_\mathrm{min}$ (light dashed curve),
$2.5\mathcal P_\mathrm{max}$ (heavy dot--dashed curve),
$2.5\mathcal P_\mathrm{min}$ (light dot--dashed curve).}\label{f6}
\end{figure}

In Section~\ref{system}, we assumed that the gravitational acceleration of plasma should
not be taken in to account in our consideration. Now we can substantiate this
assumption knowing that large velocities are achieved (Figure~\ref{f2}). Even in
the extreme case of free fall, in a time interval of $\Delta t \sim 5.3$~seconds, a
plasma parcel would gain a velocity increment of $\Delta v \sim g\Delta t \sim
1.45$~km\,s$^{-1}$, which is vanishingly small compared to the typical plasma
velocities, not to mention the burst velocities. (Including this
estimate was kindly suggested to us by the referee.)

\subsection{The Increase of the Magnetic-Pressure Gradient as the Producer of the Velocity Burst}
The idea of the physical mechanism of the velocity burst can be formed if we
consider the behaviour of other quantities during this burst. Only the region
shown in Figure~\ref{f5}, right, will be considered in what follows. The
subscripts $\mathrm{max}$ and $\mathrm{min}$ will refer to maximum and
minimum values of a quantity at hand over this very region.

\begin{figure}
  \centerline{\includegraphics[width=0.5\textwidth,clip=]{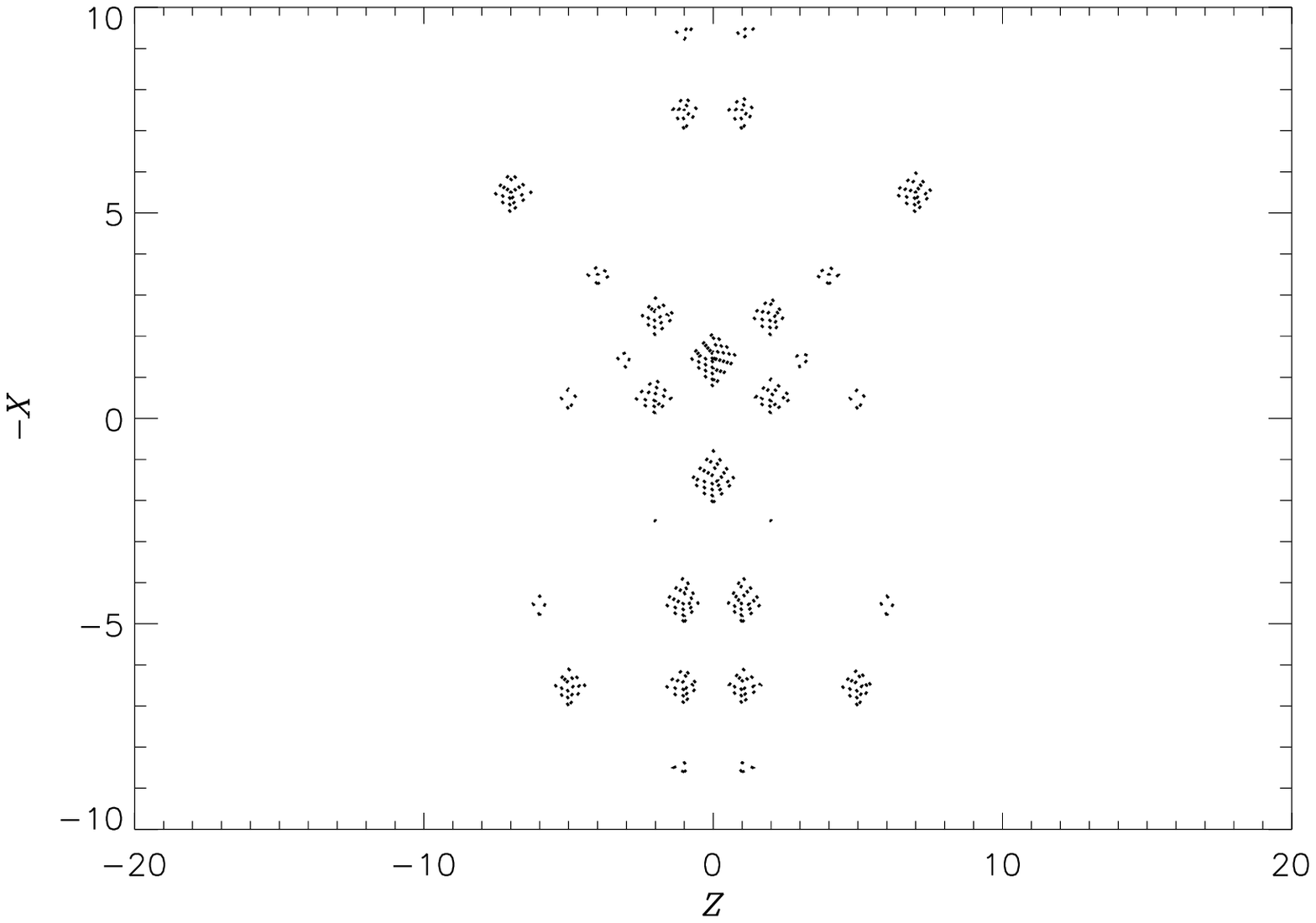}
               \includegraphics[width=0.5\textwidth,clip=]{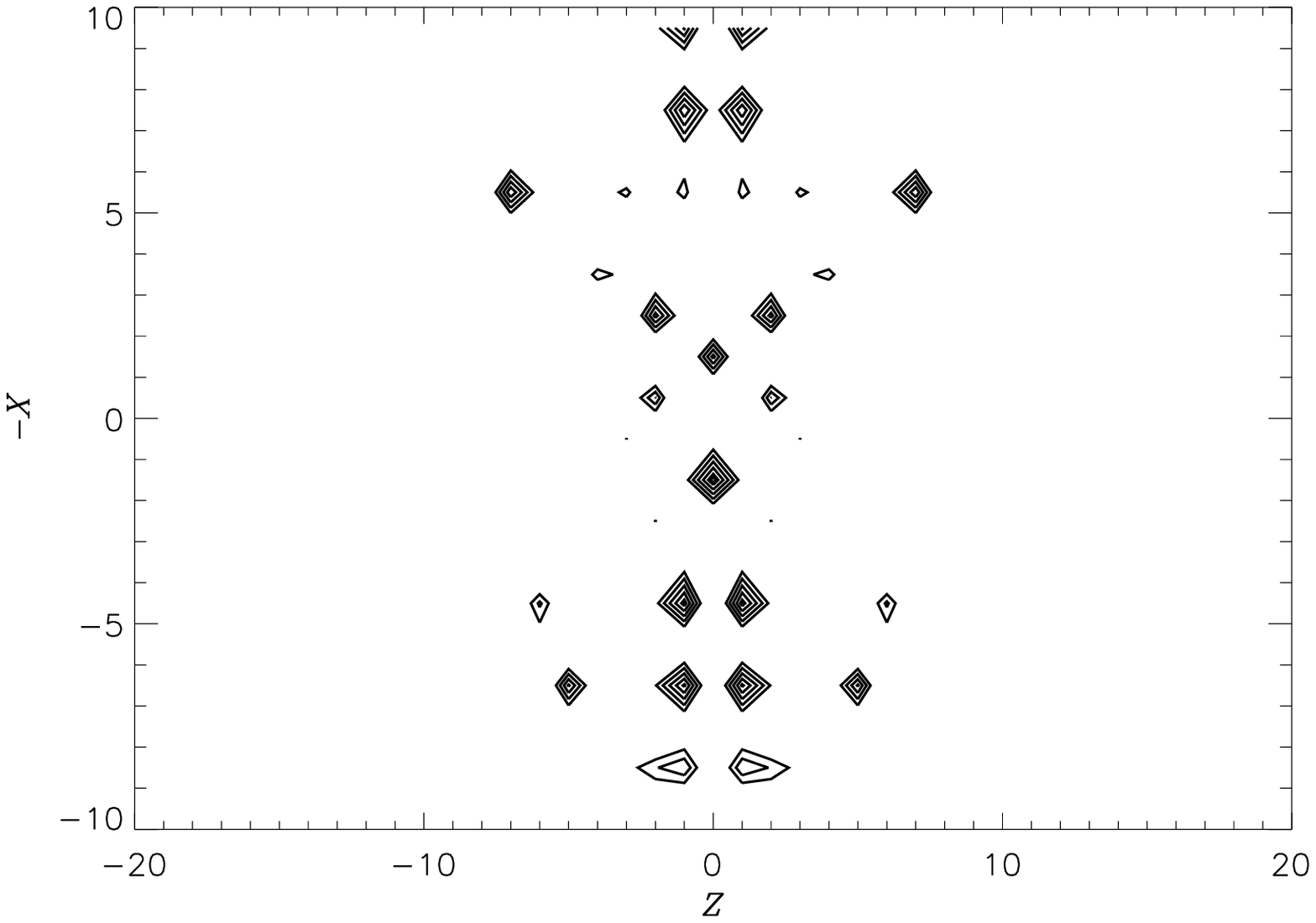}
              }
\caption{Maps of the total pressure (left) and magnetic pressure (right) at an
intermediate development stage of the burst ($t=0.2648$). Areas where the pressure
exceeds its mean value over the region are hatched.}\label{press}
\end{figure}

\begin{figure}
  \centerline{\includegraphics[width=0.5\textwidth,clip=]{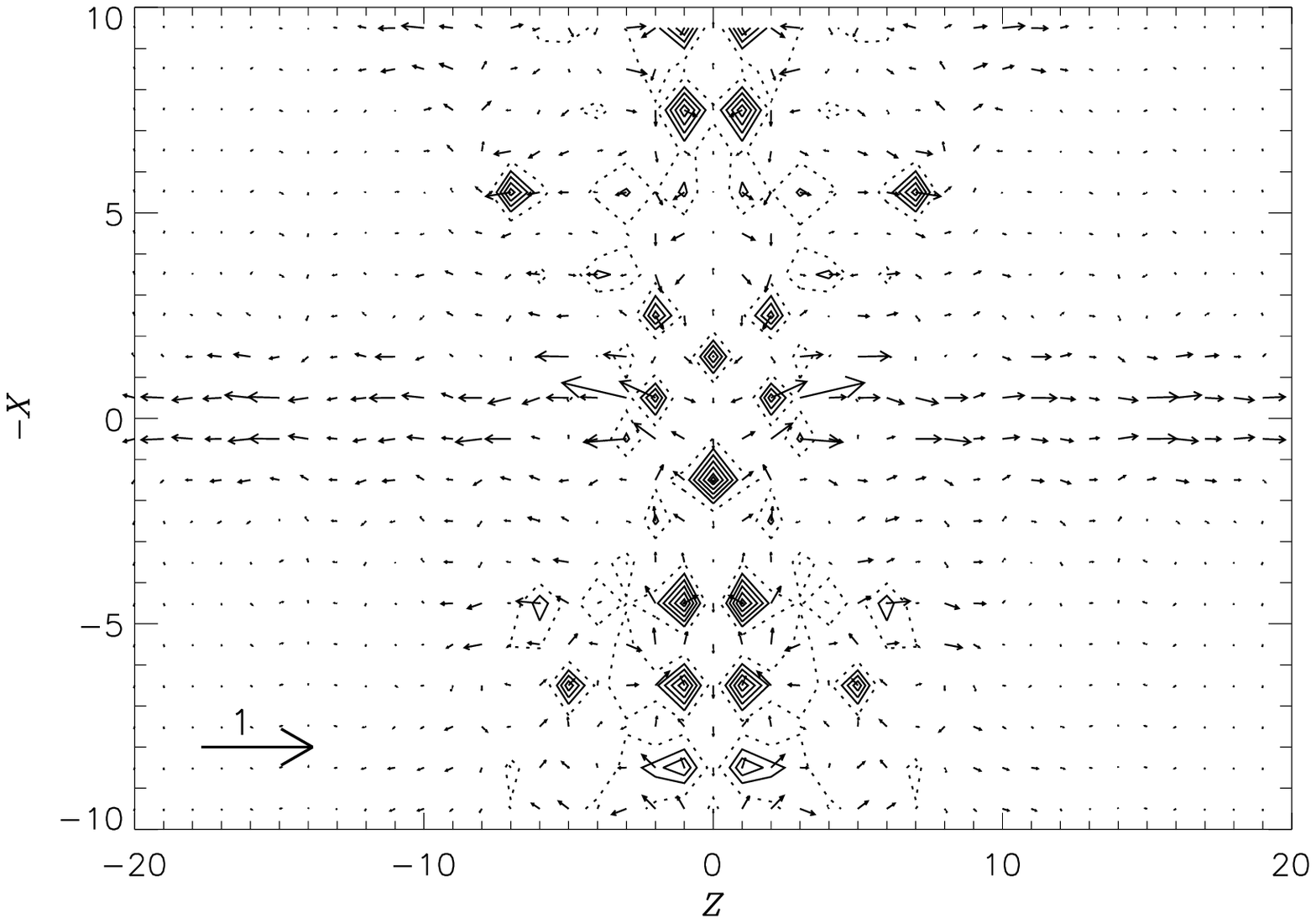}
               \includegraphics[width=0.5\textwidth,clip=]{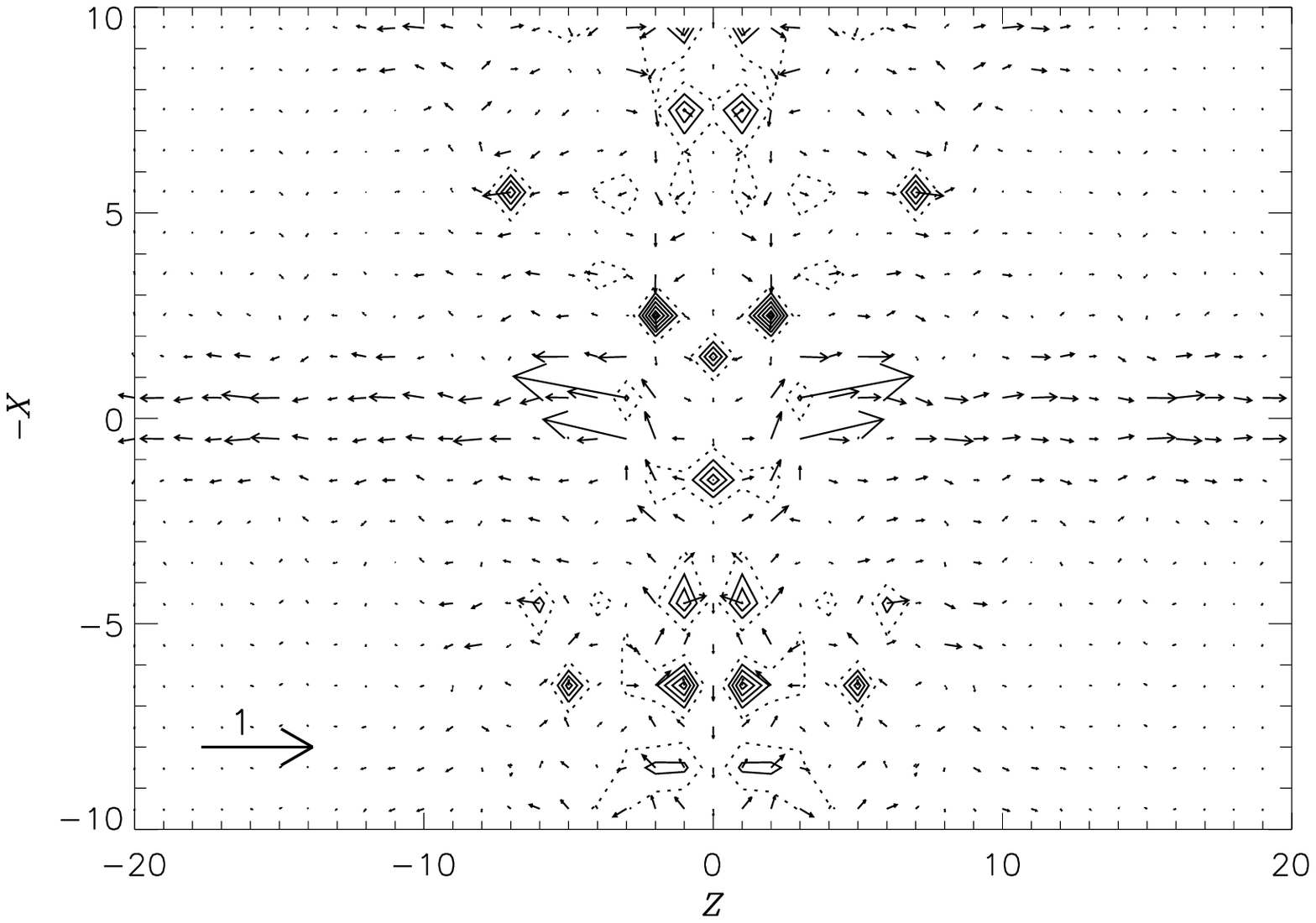}
              }
\caption{Normalised velocity $\mathbfit v_N$ and contours of the magnetic pressure for an early
($t=0.2640$, left) and the highest ($t=0.2656$, right) development stages of the burst.
The dotted curve corresponds to a level of $(1/3)[(B^2/2)_\mathrm{max}-(B^2/2)_\mathrm{min}]$.
The unit-length arrow is shown near the lower left corner of each panel; it corresponds to the
longest arrow in Figure~\ref{f5}, right.}
\label{magprar}
\end{figure}

Note that $B$, $\rho$, $\mathcal P$ and $T$ vary in the burst area not so
strongly as $v$ does (Figure~\ref{f6}). However, according to Equation
(\ref{19}), the velocity variations are directly controlled by the gradient of
the full pressure rather than the above-mentioned quantities. Note that the
elevated-gradient areas of the magnetic and the full pressure are virtually
coincident (Figure~\ref{press}), the former giving a dominant contribution to
the latter. Consider how the magnetic-pressure gradient is related to the
velocity (Figure~\ref{magprar}). Indeed, the velocity vector is aligned with
the gradient of the magnetic field pressure wherever both are graphically
representable, although this is not seen everywhere. However, the gradients
cannot always be judged by the contours plotted; this is all the more so
because even large gradients can be handled by our code using generalised
functions.

The velocity-burst mechanism could likely be described as follows. In such an
impulsive process, the velocity increase achieved over a time interval [$\delta
t$] is
\begin{equation}\label{grad}
\delta v \approx \delta t\frac{1}{\rho}\left|\nabla \frac{B^2}{2}\right|.
\end{equation}
Generally, $|\nabla B^2/2|$ can reach large values, and our
technique makes it possible to simulate cases with large
gradients.

\begin{figure}
   \centerline{\hspace*{0.015\textwidth}
               \includegraphics[width=0.85\textwidth,clip=]{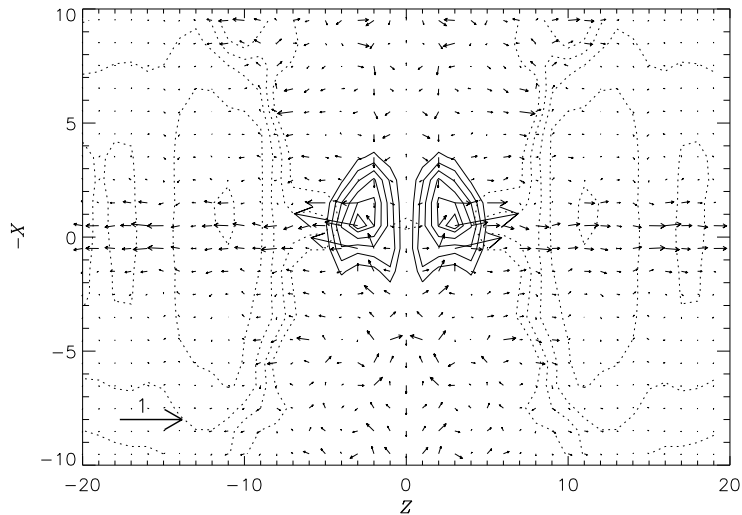}
               }
\caption{Normalised velocity $\mathbfit v_N$, temperature $T$(solid curves) and
density $\rho$ (dotted curves)
for the highest stage of burst development ($t=0.2656$).
Contour levels from 1.456 to the peak value with an increment of 0.076 are used for $T$; from 0.736 with
an increment  of 0.086, for $\rho$. The unit-length arrow is shown near the lower left corner;
it corresponds to the longest arrow in Figure~\ref{f5}, right. It can be seen that the
temperature reaches its maximum where the density is
reduced.}\label{RT}
\end{figure}

\begin{figure}
  \centerline{\includegraphics[width=0.5\textwidth,clip=]{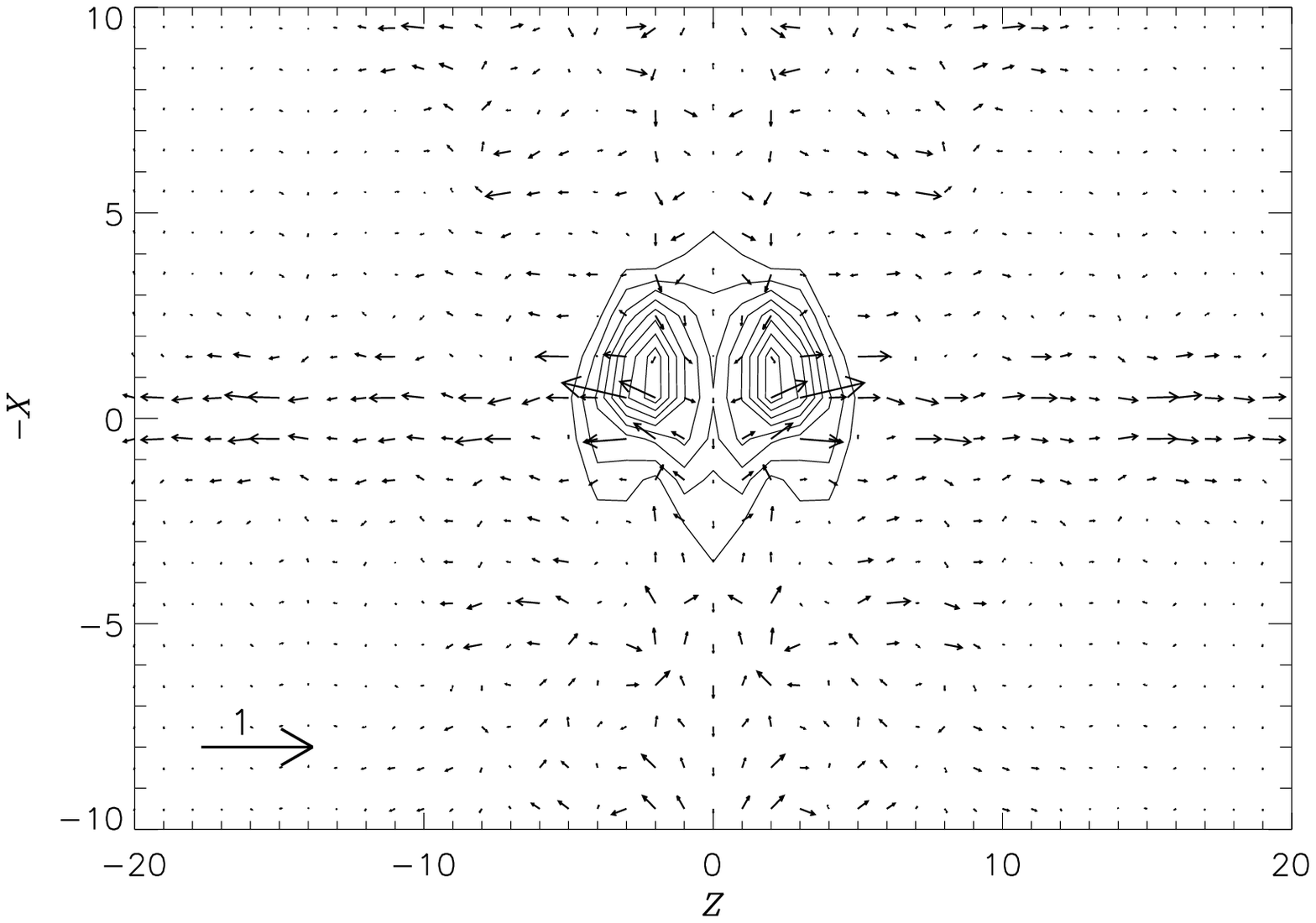}
               \includegraphics[width=0.5\textwidth,clip=]{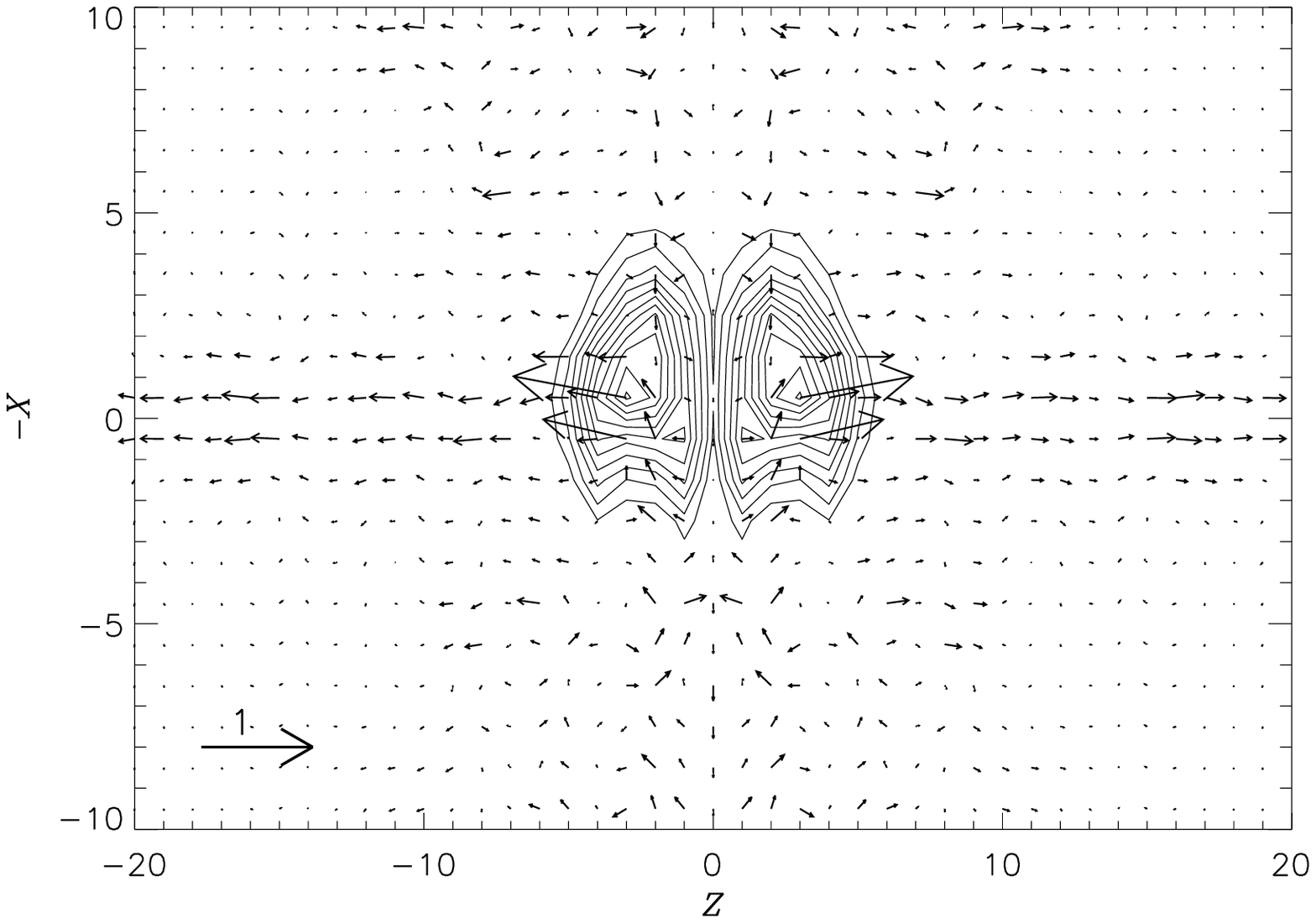}
              }
\caption{Normalised velocity $\mathbfit v_N$ and temperature for an early
($t=0.2640$, left) and the highest ($t=0.2656$, right) development stages of the burst.
The lowest contour level 1.28 and the contour increment  0.048
are the same in both panels. The unit-length arrow is shown near the lower left corner
of each panel; it corresponds to the longest arrow in Figure~\ref{f5}, right.}\label{temperat}
\end{figure}

A link between velocity increases and the presence of large magnetic-field
gradients can be revealed by an analysis of temperature variations in the burst
area. The temperature is increasing during the burst (Figure~\ref{f6}); this
occurs where the density values are reduced and the temperature increase is
therefore not an adiabatic effect of gas compression (Figure~\ref{RT}).
Actually, the plasma is heated by electric currents, and the temperature is the
higher, the larger the current density. In our 2D geometry, the current-density
magnitude is proportional to $|\nabla B|$. A comparison of the
temperature--velocity maps (Figure~\ref{temperat}) for the initial stage of the
process and for a well-developed burst shows that the longest arrows issue from
the areas of increased temperature, i.e. of increased gradients of $B$.

\begin{figure}
   \centerline{\includegraphics[width=0.5\textwidth,bb=50 0 504 360,clip=]{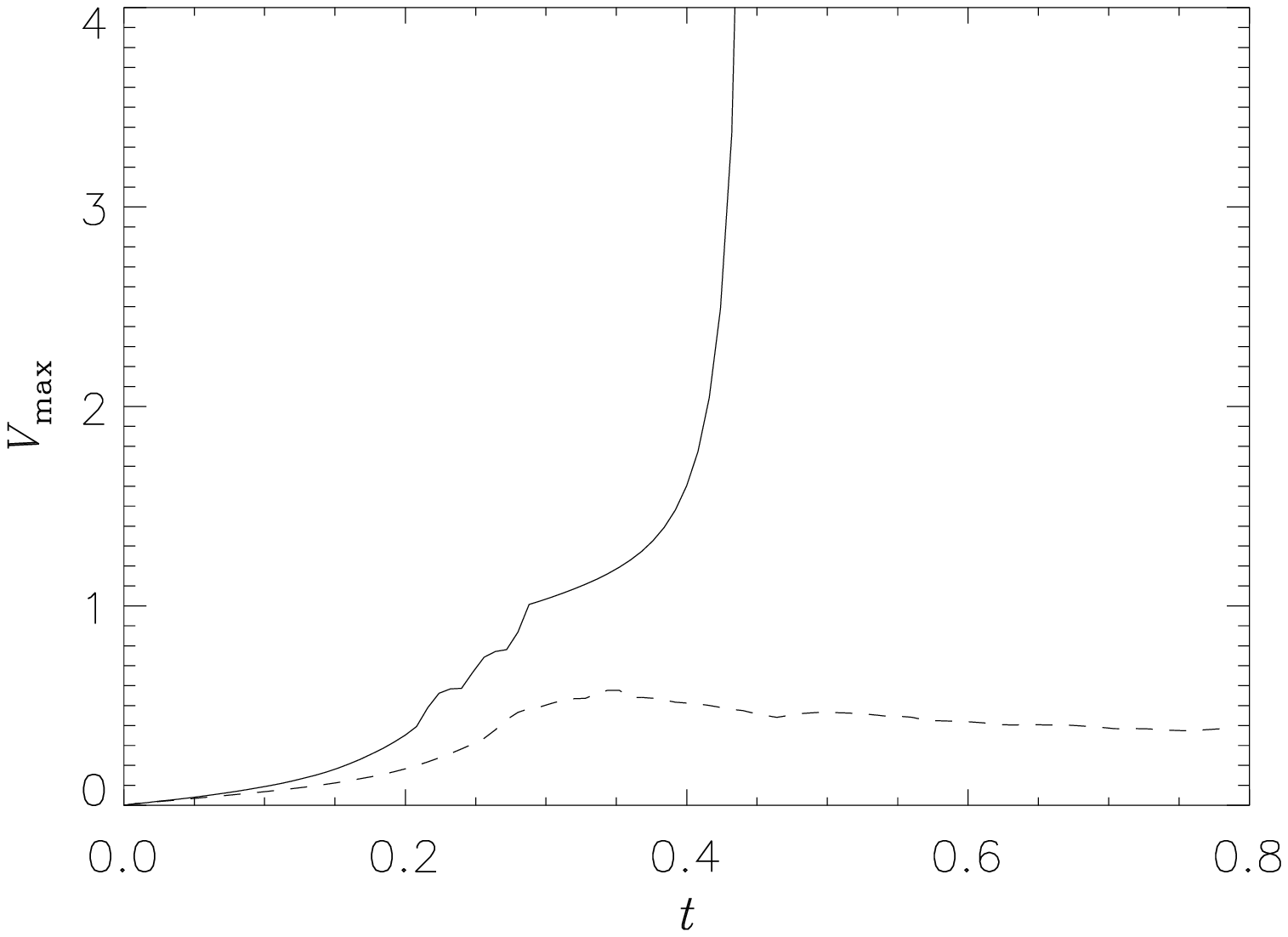}}
\caption{The maximum absolute velocity over the computation
domain as a function of time in the case of PIB at different
 viscosity $\mu_*=0$ (solid curve) and
$\mu_*=10^{-4}$ (dashed curve).}\label{f11}
\end{figure}

The velocity burst revealed here is accompanied by large
velocity gradients; therefore, this burst should be affected by
viscous forces if $M\neq 0$ is taken into account in Equation (\ref{19}).
To verify our suggestion that the velocity-burst mechanism
stems from Equation (\ref{grad}), we have carried out calculations for
the case of $M\neq 0$ assuming $\mu_*$ value exaggerated by
a factor of 20 compared with the actual viscosity calculated
according to Equation (\ref{mu}) for the considered solar-plasma layer.
The presence of such a viscosity was even able to suppress the
dramatic growth of $v_\mathrm{max}$ at the final development
stage of the pinch instability. The evaluation of the specific
peak value of $v_\mathrm{max}$ requires taking into account the
dependence of viscosity on the local temperature and using an
accurate expression for the viscous-stress tensor.

\section{Discussion}\label{discuss}
\subsection{The Onset of the Sausage Instability Against the Background of Interfering Waves}\label{interfer}
\begin{figure}
  \centerline{\includegraphics[width=0.5\textwidth,clip=]{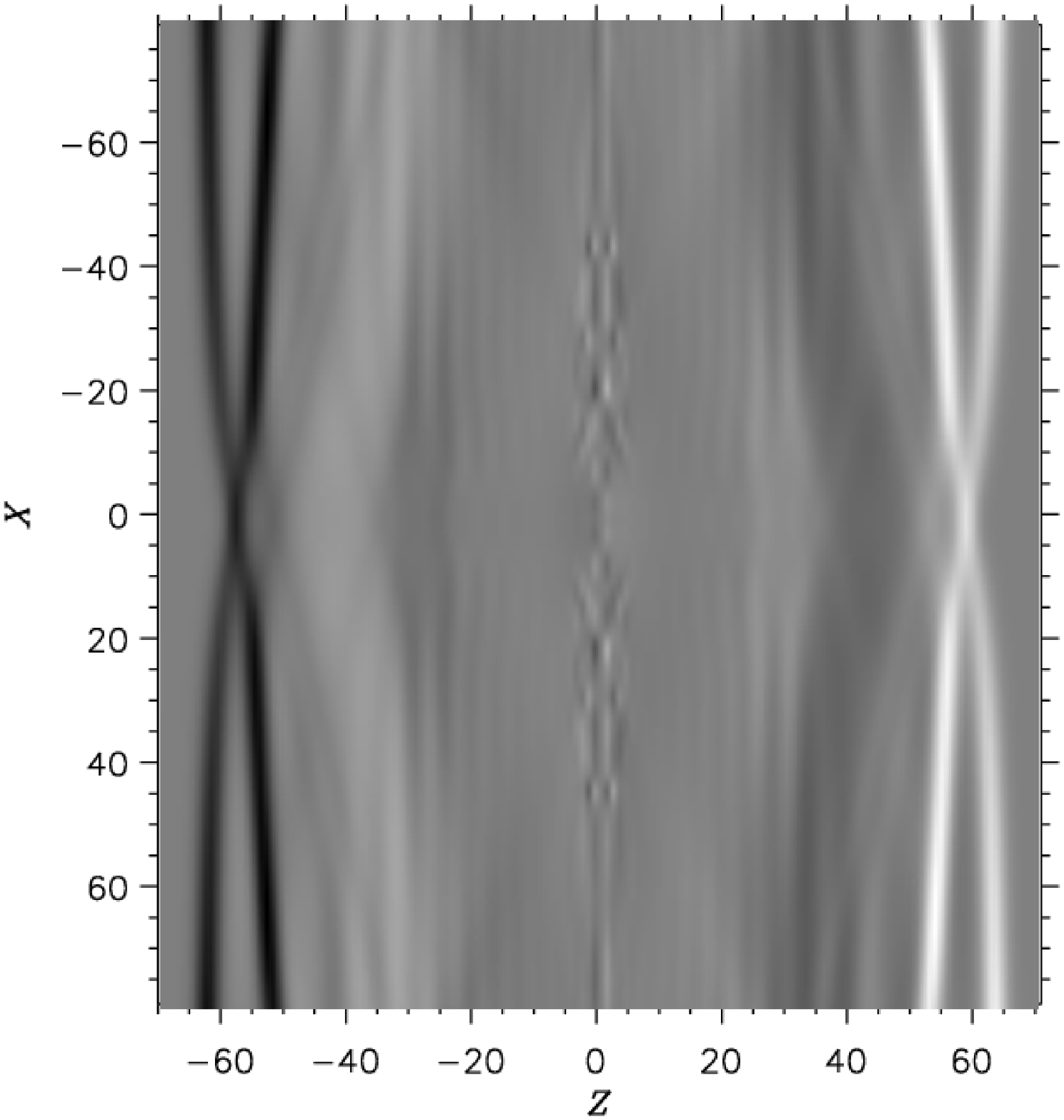}
               \includegraphics[width=0.5\textwidth,clip=]{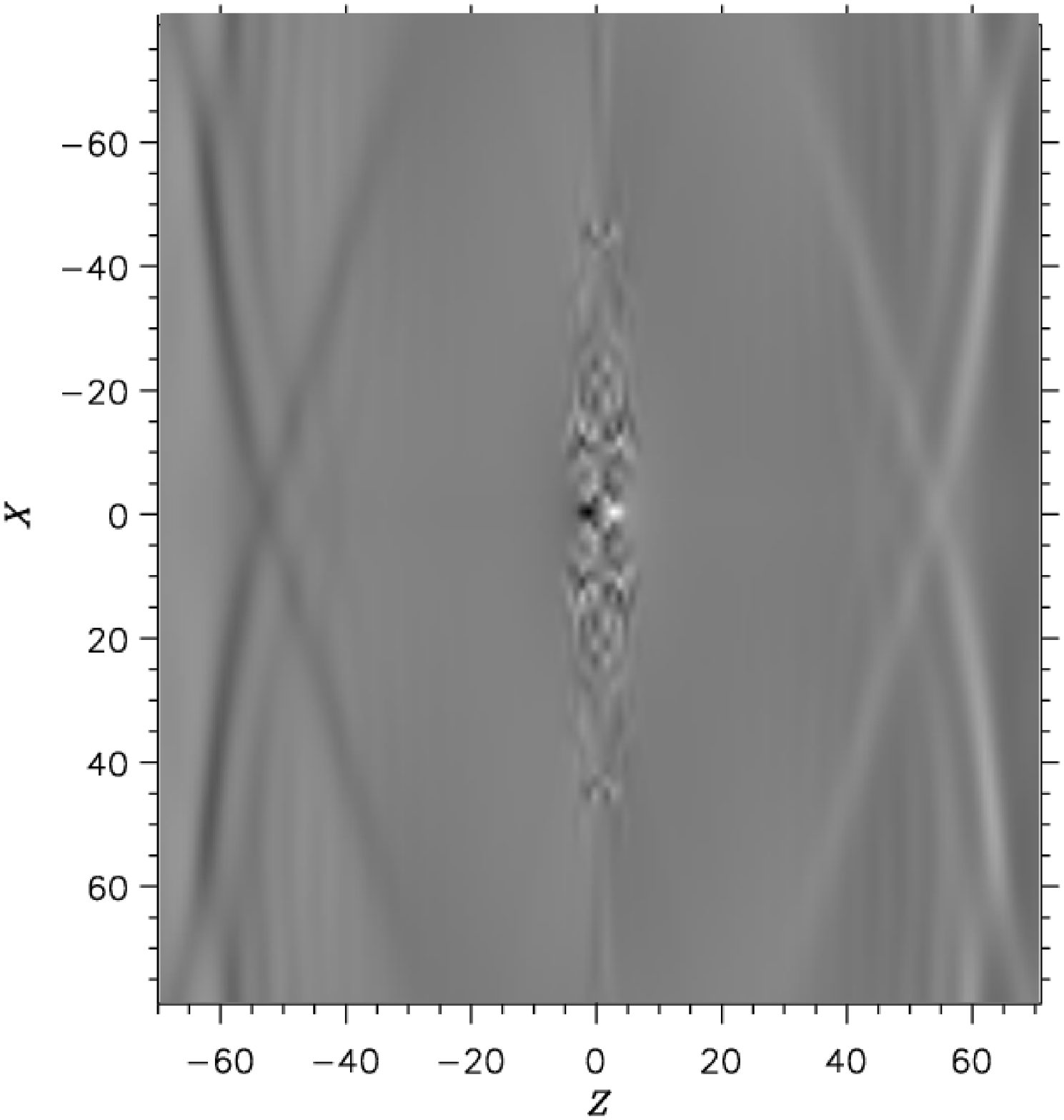}
              }
\caption{The maximum value of the velocity component along the contact zone over the computation
domain in the case of PIB $\beta_{0*}=1.5$
at $t=0.08$ (left) and $t=0.136$ (right). The sausage instability starts developing
amidst primary waves.}\label{f12}
\end{figure}

Thus, our numerical simulations reveal a velocity burst, which
is observed only at the
initial conditions of the PIB type, the case
 physically corresponding to the penetration of a new,
unbalanced magnetic field to the area of pre-existing field,
which has already reached pressure balance (Section~\ref{initial}).
This impulsive phenomenon
 is given by the spontaneously developing sausage
instability in the course of the general plasma and magnetic-field evolution
(see Section~\ref{our2}). At some time, the  instability begins developing
in the central part of the configuration amidst the primary waves produced by
the unbalanced  $\mathbfit B$ (Figure~\ref{f12}). Ultimately, it
leads to the divergence of the solution, which was predicted analytically.

The rate of the instability development should depend on the spatial
distribution of the quantities at the onset of instability (as  the
growth rate of the amplitude depends on the perturbation wavevector in equilibrium
configurations according to the linear theory of the sausage instability).
Therefore, the particular interference pattern of primary waves at the time of
onset should affect the particular scenario of instability development and,
eventually, the time of divergence.

This obvious feature of the solutions should be kept in mind comparing the solutions
for close values of the external parameters of the problem, e.g., for
$\beta_{0*} = 1.5$ and $1.6$ (Figures~\ref{f2}a,b). Although processes of the type ii) giving rise to
primary waves (see Section~\ref{our2}) differ very little, stronger magnetic
fields ensure greater amplitudes and steeper profiles of waves; therefore, the
wavevectors of the interference patterns is highly sensitive to the parameter
$\beta_{0*}$ variations. This
results in different temporal variations of the maximum velocity seen in the
\textit{white} regions of Figures~\ref{f2}a and \ref{f2}b (note that, at
different times, the maximum velocity is generally reached at different points,
directly depending on the temporal variation of the interference pattern).
At later times, when the
disturbance related to the sausage instability becomes strong,
 the differences in the behaviour of the curves become greater
 (the \textit{shaded} regions of Figures~\ref{f2}a and \ref{f2}b).
Thus, the manifestations of the velocity burst depend on the fine details
of the early development stage of the instability.

\subsection{Velocity Burst and the Problem of Dynamic Chromospheric Energisation}
In our initial-value problem, the plasma is initially
motionless and has a constant temperature of 50,000~K
throughout the plasma volume. Further, the magnetic-field
energy transforms into the thermal and kinetic energy of the
plasma because of the development of a pinch with a
sausage instability.

For example, in the numerical simulations at $\beta_{0*}=1.6$ in the
absence of gas viscosity,
 the plasma temperature increases by a factor of
two, whereas the kinetic temperature of the proton
component reaches a peak value of 0.511~MK at undimensional $t=0.2656$, or an average
particle energy of 44~eV, and substantially
exceeds it at later times during the final velocity burst
(Figure~\ref{f2}b; we note that the exact solution goes to
infinity\,--\,see Section~\ref{analyt}). We see that impulsive processes
can manifest itself against the
background of a gradual development of the pattern.
Specifically, intense plasma streams emerge
in a temporal interval of about 0.1~second or less near the centre of the magnetic
configuration. Such a velocity burst can physically
be interpreted as an event of suprathermal-ion generation.  This is
indicative of the dynamic energisation of a parcel of chromospheric plasma to
coronal temperatures and, if the final stage is achieved, of the
increase in the kinetic
temperature to high values, even those observed in flares.

Note that magnetic reconnection is not present in our solution: it would
generate plasma flows varying \textit{along} the magnetic-field line,
which is forbidden by the chosen 2D geometry without variations of the physical quantities
in the direction of $\mathbfit B$. Therefore, under the conditions
of the upper chromosphere, the pinch-sausage  effect
can be considered a possible alternative to the magnetic-reconnection process
  as the producer of flares. However, it can develop only in the presence of the ``old''
  magnetic field, which has already reached pressure balance (we recall that, in Figure~\ref{f2},
  the curves for PIB case, in contrast to NIB case, demonstrate the velocity burst; see also Appendix C).

To describe the velocity burst more accurately,
  the equation system
  used here should be generalised by adding
  more physical processes. Thus,
our simulations suggest that a finite gas viscosity can reduce
the kinetic temperature of the fast streams. There are a number of possibilities to obtain more
 realistic solutions: introducing an adequate gas viscosity in the
system of collisional MHD equations; including
terms with the Hall-plasma-dynamics parameter $\xi$, i.e. considering a
conductivity tensor with $\sigma_\mathrm{H}$ and $\sigma_\mathrm{P}$ present;
 and proceeding to
three-dimensional simulations (which, in particular, could make it possible to
consider also magnetic reconnection). This work is
currently underway.

\subsection{Velocity Burst and the Ion Beams}
 The suprathermal ions (protons) revealed in our simulations are related to a
phenomenon known from laboratory nuclear-fusion experiments probably.
Specifically, neutrons were produced in Z-pinch plasma by relatively
small quantities of ``beam'' deuterium ions accelerated in the
direction of the current to energies of 50\,--\,200~keV due to the
development of the ``sausage'' $m=0$ instability of the pinch.
Colliding with the deuterium ions whose temperature was much lower,
the beam ions produced fusion neutrons, which were therefore not
thermonuclear \citep{Z-P2005}. The origin of this phenomenon due to
the nonlinear evolution of the $m=0$ instability of a stagnated
compressional Z-pinch  is still not fully understood. Some of the
proposed mechanisms are fluid-like and others are kinetic in nature.
From the  MHD standpoint, ion jets in both directions should occur
equally from the ``necked'' region. Presumably, it requires
introduction of the Hall term and/or, the finite ion Larmor radius
terms in the stress tensor to break this symmetry and allow a beam in
the direction of the current to form \citep{rev}.

Some results of our numerical experiment with MHD hydrogen
proton plasma show evidence of their correspondence to this
situation. In the PIB case, where stagnation is initially
present, a nonlinear state of the pinch process gives rise to
opposite current-aligned ion jets of short duration in the
centre of the magnetic configuration
(Figures~\ref{f2}~and~\ref{f5}). The energies of ions accelerated during the final
velocity burst would be comparable with the energy value
indicated above for the deuterium ions obtained in
laboratories. It is worth adding that the simulated pair of
suprathermal-ion jets could actually produce an ion beam
accelerated in the direction of the current. Indeed,
suprathermal ions experience less frequent collisions than the
thermal particles of the plasma and are more directly
controlled by the electromagnetic field. Near the symmetry axis
along which they move, the magnetic field is weak. The electric
field should accelerate the ions of one jet and decelerate the
ions of the oppositely directed jet. Therefore, only one ion
beam will be produced.

\section{Conclusion}
Contacted oppositely directed horizontal magnetic fields
are not rare in the upper chromosphere, where the plasma is fully ionised.
 Therefore, the phenomena described by our numerical
solutions should not be unusual.

We have traced the coevolution of the plasma and magnetic field from the origin of the steadying waves
to the final stage when a velocity burst occurs. The temperature of the chromospheric plasma
increases by two to three times in the course of the evolution.
The velocity burst is due to the sausage instability
 that develops against the background of the steadying waves, turbulence, frontal zones, \textit{etc.}
As a result, fast streams suddenly appear at the centre
of the  magnetic configuration.
 The kinetic temperature corresponding
  to such fast streams can  approach the values recorded
  in the corona and, likely, in flares.
  So, the pinch-sausage  effect can be a possible energiser
  of the upper chromosphere and an alternative to the magnetic-reconnection process
  in producing  flares.

We have numerically solved the standard self-consistent MHD system of equations.
This made it possible to relate the plasma--magnetic--field coevolution
to the initial values of the basic MHD   quantities without invoking  ad-hoc non-observable parameters.
The standard form of the MHD system allows us to compare our results for the solar plasma
with those for the laboratory plasma, and
 we note that the velocity burst revealed in our simulations resembles
  the sudden emergence  of high-speed proton  beams in laboratory Z-pinch devices.

\appendix
\section{Material Properties of the Medium}

\textit{Magnetic diffusivity}. Following \inlinecite{spitzer} and
\inlinecite{Priest}, we define the conductivity [$\sigma$] of a collisional,
fully ionised hydrogen plasma as
\begin{equation}\label{sigma}
\sigma=\frac{e^2N_{\mathrm e}\tau_{\mathrm e}}{m_{\mathrm e}},
\end{equation}
where  $N_\mathrm e$ and $\tau_e$ are the local electron concentration and
electron--ion collision time. In a fully ionised gas there is some uncertainty
 as to the appropriate value of $\tau_{\mathrm e}$ to use in this equation \citep{spitzer}.
 We choose $\tau_{\mathrm e}$ in accordance with formula (2.5e) of \inlinecite{Brag}, which
 coincides with  $\tau_{\mathrm e}$ in \S\ 5.7 of \inlinecite{balescu}. [Thus,
 our $\sigma$ is equal to that used by \inlinecite{BM} and \inlinecite{Brushl1989}. It is also used as a basis
for the generalisation of the form of conductivity to the case of plasma with
neutral particles \citep{arber2006}.] Explicitly writing $\tau_e$
 yields $\Theta$ in the form
 Equation (\ref{TetK}), where
\begin{equation}\label{teta}\theta_*=\sqrt{\frac{m_\mathrm em_\mathrm
i}{2\pi}}\frac{c^2}H\frac 1{(kT_*)^2}\frac{\Lambda e^2}{0.75}
\end{equation}
($\Lambda$ is the Coulomb logarithm, which we set here equal to
20).

In a quasi-neutral plasma, $\sigma \propto N_\mathrm e\tau_\mathrm e$ is a
function of only the local temperature.

\textit{Thermal conductivity and gas-dynamic viscosity coefficient.}

  The thermal conductivity of a collisional
plasma differs substantially between the
region of magnetised plasma and the region of zero magnetic
field. In the (dimensional) heat-transfer
equation \citep{Brag},   we use the quantity corresponding
to the case of weakly magnetised plasma (where, in the commonly
accepted notation, $\kappa_\perp \sim \kappa_\perp^e \sim
\kappa_\parallel^e$).  Once the
nondimensionalisation of the equations is accomplished, this
coefficient, appearing in the expression for Equation (\ref{KTet}),
 determines
\begin{eqnarray}
\label{kapp}\kappa_*=\frac{k_s{T_*}^2{\sqrt{m_\mathrm
i}}}{32Hk^{3/2}N_*},\\
\label{kappas}k_s=\frac{3.16\cdot{0.75}}{\Lambda\sqrt{2\pi}}\frac
{k^{7/2}}{\sqrt{m_\mathrm e}\cdot{e^4}},
\end{eqnarray}
(note that the so-called Spitzer thermal conductivity [$k_s$] has dimensions of
$\mathrm{erg}\,\mathrm{s}^{-1}\,\mathrm{cm}^{-1}\, \mathrm{deg}^{-7/2}$). This
substantially overstates the thermal conductivity in the region of the strong
magnetic field; however, as our results indicate \citep{A-K-SAO}, major
 thermal
 changes occur in
the region of weak magnetic field $x=x_\mathrm c$.

The expression for the kinematic-viscosity coefficient of plasma with a
magnetic field is very complicated \citep{Brag}. Instead, even in the case of a
magnetic field present, \inlinecite{Priest} and
 \inlinecite{spitzer} consider the coefficient
of plasma without magnetic field
\begin{equation}\label{mun}
M_0=2.21\times{10}^{-15} \Lambda^{-1}(T_\mathrm i/\mathrm K)^{5/2}
\mathrm g \cdot {\mathrm c\mathrm m}^{-1}\cdot{\mathrm s}^{-1},
\end{equation}
where $T_{\mathrm i}$ is the ion temperature in $\mathrm K$ (since the
gas-dynamic viscosity of plasma is determined by ions,  it is worth recalling
 that the ion temperature is half the plasma
 temperature under consideration).
 That practice is
appropriate for use in our investigation, since, in the
magnetic fields studied with the pinch effect, the most interesting
phenomena arise in the neighbourhood of the neutral line where
magnetic field  is not great. So  we obtain $M$
in Equation (\ref{Mvis}) with the dimensionless parameter [$\mu_*$] determined
by
\begin{equation}\label{mu}
\mu_*=\frac {0.215} \Lambda \frac{T_*/(1\ \mathrm K)}{N_* \times (1\ \mathrm {cm}^3 )}.
\end{equation}
The present study addresses the qualitative role of viscosity;
 therefore, the simplest form of the viscous force is used
 in Equation (\ref{19}).

\section{Initial Magnetic-Field Distribution}

The structure of the magnetic field at the initial time is determined
by the expression
\begin{eqnarray}\label{nachvid}
B(x,z)|_{t=0}&=&D(x,z)+d(x,z),\\
\label{forbeta}D&=&\left[1+\zeta\exp(-\zeta)\right]\tanh[q(x-x_\mathrm{c})],\quad
\zeta\equiv(z-z_\mathrm{c})^2/z_k^2,
\\
\label{outbeta}d&=&D\frac{s(x-x_\mathrm{c})^2\exp(-1000\zeta)}
{{[x_s^2+(x-x_\mathrm{c})^2]}^2[1+s_1(z-z_\mathrm{c})^2]},
\end{eqnarray}
where $q=130$, $x_s=0.031$, $x_\mathrm{c}=x_\mathrm{max}/2$,
$z_\mathrm{c}=z_\mathrm{max}/2.$

Plasma with zero gas-dynamic viscosity was simulated in a computational
domain with $x_\mathrm{max}=0.207$, $z_\mathrm{max}=2.8$ at
$s_1=1.76$, $z_k=0.17$, $s=0.01$. With this choice of
constants, the term $|D|$ in Equation (\ref{nachvid}) is approximately
half of the value of $|d|$. For a viscous plasma, the calculations
were carried out in a computational domain with
$x_\mathrm{max}=0.207$ and $z_\mathrm{max}=0.639$ at $s_1=0$,
$z_k=0.025$, $s=0.00025$. Therefore, we get
$|D|\approx |d|$.

\section{Initial Pressure and Density in the PIB Case}

Recall that $\rho(x,z)|_{t=0} =1$ outside the region occupied by the magnetic
field. Assume that, in the region of the magnetic field Equation
(\ref{nachvid}), only the component $D$ is in equilibrioum with the gas
pressure:
\begin{equation}\label{Pnach}
{\cal P}(x,z)|_{t=0}+ {D(x,z)}^2/2=\beta_{0*}/2.
\end{equation}
Taking into account Equation (\ref{sost}) leads to the following initial
conditions for pressure and density:
\begin{eqnarray}\label{P2}
{\cal P}(x,z)|_{t=0}&=&\beta_{0*}/2-{D(x,z)}^2/2,\\
\label{Rho2}\rho|_{t=0}&=&1-\frac{{D(x,z)}^2}{\beta_{0*}}.
\end{eqnarray}

\section{The Applicability of the Considered Initial-Value Problem to the
Case of $\beta_{0*}=1.5-2.3$}

Weak magnetic field evolves slowly and requires computations over long periods
to reach the divergence of the
solution (described in Section \ref{analyt}). This leads to accumulation of
errors. Therefore, we choose the characteristic $B$ value to be as large as $B_0=0.33\
\mathrm G$.

Although this  $B_0$ corresponds to characteristic values $\omega_{e0}\tau_{e*}
\gg 1$, we nevertheless use it to  solve numerically the initial-value
problem\, -- \, Equations (\ref{19})\, --\, (\ref{teplo}),
 (\ref{VTnach}), (\ref{nachvid}) and either
(\ref{Rho1}), (\ref{P1}) or (\ref{P2}), (\ref{Rho2})\, --\, with isotropic
$\Theta , K, M$, intending to supplement the system, at a later time, with
terms including the parameter $\xi$, or, in other words, to automatically take
into account the anisotropy of the local coefficients in the areas where this
anisotropy is substantial (see footnote in Section \ref{problem}).

However, in the  magnetic configuration considered (Figure~\ref{f1}a), the most
important phenomena develop near the  line of contact of opposite polarities
$x=x_\mathrm c$, where magnetic field vanishes and the local
coefficients are therefore isotropic (see Section \ref{analyt}). Thus, the results
obtained for the given $B_0$ appear to adequately reflect the reality. (We can
also note that our tentative calculations with $\xi \neq 0$ for the same
magnetic configuration and initial conditions show virtualally no difference with
the case of $\xi=0$ presented here; only a more rapid evolution is observed.)

\begin{acks}
We cordially thank A.V. Getling for carefully considering the ideas of this
study and useful recommendations. We are also grateful to the referee for
discussions. This work was supported by the Russian Foundation for Basic
Research (project no. 12-02-00792-a).

\end{acks}
\section*{Disclosure of Potential Conflicts of Interest}
The authors declare that they have no conflicts of interest.

\bibliographystyle{spr-mp-sola}
\bibliography{bibdata}

\end{article}
\end{document}